\documentclass[10pt]{iopart}

\usepackage{iopams}  
\usepackage{isomath}  
\usepackage{dsfont}  
\usepackage{graphicx}
\usepackage{bm}
\usepackage{amssymb}
\usepackage{mathrsfs}
\usepackage{color}
\usepackage{cancel}
\usepackage{tikz} 
\usepackage{tkz-euclide}
\usepackage{tabularx}
\usepackage{pgfplots}
\pgfplotsset{compat=1.7}
\usepackage{enumitem}
\usepgfplotslibrary{patchplots}
\usepgfplotslibrary{fillbetween}
\usepackage{bibentry}
\usetikzlibrary{snakes}
\usetikzlibrary{math} 
\usepackage{caption}
\usepackage{subcaption}
\usepackage{wasysym}

\begin{document}
	
	\title{Effect of mismatch on Doppler backscattering in MAST and MAST-U spherical-tokamak plasmas} 
	
	\author{
		V.H. Hall-Chen$^{1,2,3}$, 
		F.I. Parra$^{4}$, 
		J.C. Hillesheim$^{5}$,
		J. Ruiz Ruiz$^{1}$, 
		N.A. Crocker$^{6}$, 
		P. Shi$^{2}$, 
		H.S. Chu$^{3}$,
		S.J. Freethy$^{2}$, 
		L.A. Kogan$^{2}$,
		W.A. Peebles$^{6}$, 
		Q.T. Pratt$^{6}$, 
		T.L. Rhodes$^{6}$, 
		K. Ronald$^{7}$, 
		R. Scannell$^{2}$,
		D.C. Speirs$^{7}$, 
		S. Storment$^{6}$,
		J. Trisno$^{3}$
	} 
	
	\address{
		$^1$ Rudolf Peierls Centre for Theoretical Physics, University of Oxford, Oxford OX1 3PU, UK\\
		$^2$ UKAEA/CCFE, Culham Science Centre, Abingdon, Oxon, OX14 3DB, UK\\
		$^3$ IHPC, Agency for Science, Technology and Research (A*STAR), Singapore, Singapore 138632, Singapore\\
		$^4$ Princeton Plasma Physics Laboratory, Princeton, NJ 08540, USA\\
		$^5$ Commonwealth Fusion Systems, Cambridge, MA, USA \\
		$^6$ Department of Physics and Astronomy, University of California, Los Angeles, CA 90095, USA \\
		$^7$ Department of Physics, SUPA, University of Strathclyde, Glasgow G4 0NG, UK \\
		}
	
	\ead{valerian\_hall-chen@ihpc.a-star.edu.sg}
	\vspace{10pt}
	\begin{indented}
		\item[]October 2023
	\end{indented}
	
	\begin{abstract}
		The Doppler backscattering (DBS) diagnostic, also referred to as Doppler reflectometry, measures turbulent density fluctuations of intermediate length scales. However, when the beam's wavevector is not aligned perpendicular to the magnetic field, the backscattered power is attenuated. In previous work, we used beam tracing and reciprocity to derive this mismatch attenuation quantitatively. This theoretical framework was implemented in a predictive code, Scotty, and validated on DIII-D plasmas, where the mismatch angle at cut-off was varied systematically via toroidal steering. In this paper, we apply Scotty's predictions to the following novel regimes: spherical-tokamak plasmas, a different polarisation (O-mode), and a wider range of frequencies (30--60 GHz). Since spherical tokamaks have larger magnetic pitch angles and shears than conventional tokamaks, the former provides a good avenue for testing our ability to quantitatively calculate mismatch attenuation. For both O- and X-modes, we compared experimental data with our Scotty's predictions at multiple times during the shots, showing consistent good agreement, an improvement over earlier work, where only one time per shot was studied. Finally, we analysed other contributions to the backscattered signal, such as the probe beam's electric field and the turbulence spectrum. For MAST and MAST-U, we find that the effect of these contributions on measurements of mismatch attenuation is smaller than experimental uncertainty. Hence, the variation of DBS signal on toroidal steering depends predominantly on mismatch attenuation alone. The high sensitivity of DBS to mismatch indicates that DBS can potentially shed light on magnetic pitch angle in the core; as such, this paper potentially provides the physics basis for a new pitch-angle diagnostic.
		


	\end{abstract}
	
	%
	%
	%
	%
	\ioptwocol

	\section{Introduction}
	The Doppler backscattering (DBS) diagnostic, also referred to as Doppler reflectometry, measures flows and turbulent density fluctuations of intermediate length scales, typically $10 \gtrsim k_\perp \rho_i \gtrsim 1$  \cite{Hirsch:DBS:2001, Hillesheim:DBS:2009, Hillesheim:DBS_rotation:2015, Conway:DBS_Flow:2005, Conway:flows:2010, Schmitz:LH:2017, Tynan:drift_turbulence:2009, Pratt:DBS:2022}. Here $k_\perp$ is the turbulence's wavenumber and $\rho_i$ is the ion gyroradius. Additionally, the DBS diagnostic is able to measure turbulence in the core, which is challenging. It is also likely to enjoy continued usage in future fusion reactors, as it is robust enough to survive the high neutron fluxes involved \cite{Volpe:microwave:2017, Costley:new_diagnostics:2010}. Given the usefulness and versatility of DBS, it has been implemented at a wide range of tokamaks and stellarators \cite{Hennequin:DBS:2004, Happel:DBS:2009, Zhou:DBS:2013, Happel:DBS_synthetic:2017, Shi:DBS:2016, Rhodes:DBS:2016, Hu:DBS:2017, Tokuzawa:DBS_LHD:2018, Schmitz:DBS:2018, MolinaCabrera:DBS_TCV:2019, Wen:DBS:2021, Tokuzawa:DBS_LHD:2021, Carralero:DBS_JT60SA:2021, Yashin:DBS:2021, Rhodes:DBS:2022, Shi:DBS_MASTU_SWIP:2023, MolinaCabrera:DBS_ASDEXU:2023, Liu:DBS:2023}.
	
	The DBS diagnostic works as follows. A microwave beam is sent into the plasma, and the backscattered signal is measured by the same antenna that emits the probe beam. As such, the wavenumber of turbulence responsible for backscattering at any point along the beam's path is twice that of the beam's wavenumber. This is commonly known as the Bragg condition. Additionally, the wavevector of the turbulence has to be parallel to that of the beam. This condition is not always exactly met. The turbulence wavevector perpendicular to the magnetic field is much larger than that parallel to it, hence the beam wavevector has to also be perpendicular to the field. We call the angle between beam and turbulence wavevectors the mismatch angle, and the associated decrease of the signal the mismatch attenuation. The DBS signal thus depends on both density turbulent fluctuations, which one tries to measure, and mismatch attenuation. In addition to DBS, analogous wavevector mismatch is also important for the cross-polarisation scattering diagnostic \cite{Hong:CPS:2021} and the high-k scattering diagnostic \cite{Mazzucato:high_k:2006}. To avoid this problem in DBS, mismatch attenuation is typically minimised by toroidally steering the probe beams such that the mismatch angle is zero at the cut-off \cite{Rhodes:mismatch:2006, Hillesheim:DBS_MAST:2015, Damba:mismatch:2021, Hall-Chen:beam_model_DBS:2022, Damba:mismatch:2022, Hall-Chen:mismatch:2022}, from where most of the backscattered signal is expected to come. Unlike electron-cyclotron heating and current drive beams \cite{Smits:EC:1993, Cengher:EC:2016}, actively matching the polarisation ellipticity to the edge magnetic field for perfect coupling is not practised by the DBS community \cite{Hillesheim:DBS_MAST:2015, Damba:mismatch:2022, Carralero:DBS_JT60SA:2021, Shi:DBS_MASTU_SWIP:2023, Ponomarenko:DBS:2023}.
	
	However, toroidal optimisation of mismatch is more complicated in spherical tokamaks, since the magnetic field pitch angle is not only large, but also varies both temporally and spatially. Hence, if the DBS has multiple launch frequencies or when the equilibrium is changing throughout the shot, as was the case for the Mega Ampere Spherical Tokamak (MAST) \cite{Lloyd:MAST:2003}, it is not be possible to optimise steering for all frequencies or at all times, respectively \cite{Hillesheim:DBS_MAST:2015}. Even in the best of cases, most of the frequencies will have at least a significant mismatch attenuation most of the time. The need for a quantitative understanding of mismatch is more pressing, with MAST Upgrade (MAST-U) starting operations recently \cite{Harrison:MAST-U_overview:2021}, the National Spherical Torus Experiment Upgrade (NSTX-U) expected to begin operations soon \cite{Gerhardt:NSTX-U_recovery:2021}, and the Spherical Tokamak for Energy Production (STEP) \cite{BEIS:UK_strategy:2021} as well as the spherical tokamaks from private companies \cite{Gryaznevich:ST40:2017} planned for the future. MAST-U has two DBS systems, one from the Southwestern Institute of Physics (SWIP) \cite{Shi:DBS:2019, Wen:DBS:2021, Shi:DBS_MASTU_SWIP:2023} and the other from the University of California, Los Angeles (UCLA) \cite{Storment:DBS:2021, Peebles:DBS:2010, Rhodes:DBS:2022}. Data from the latter is studied in this paper. Being able to correct for mismatch attenuation in these systems will enable us to properly characterise the turbulence in spherical tokamaks, which is not only interesting for the rich physics involved but will also help us design better spherical tokamaks in the future. Moreover, recent work \cite{Damba:mismatch:2021, Damba:mismatch:2022, Hall-Chen:mismatch:2022} indicates that mismatch attenuation is significant even in conventional tokamaks; the same techniques demonstrated in this paper would also apply to them.
	
	To evaluate our understanding of mismatch attenuation and thus our ability to account for it, we study a group of repeated shots carried out at MAST and another on its upgrade, MAST-U. The first of these shot groups had previously been analysed with a heuristic estimate of the mismatch attenuation \cite{Hillesheim:DBS_MAST:2015}, and properly evaluated only at a single time \cite{Hall-Chen:beam_model_DBS:2022, Hall-Chen:mismatch:2022}.
	In this work, we compare our model with multiple O-mode channels for the first time. We then apply our model to both O-mode and X-mode data from different times within each shot group, giving further confidence to our model's ability to quantitatively predict mismatch. This analysis enables us to identify a systematic error in the Q-band system of the MAST DBS. Finally, we show how other contributions to the backscattered signal might affect our measurements of mismatch attenuation. 
	
	We begin by summarising relevant insights and the associated formalism from our previous work \cite{Hall-Chen:beam_model_DBS:2022} on the beam model of DBS in Section \ref{section:beam_model}. We then detail how experimental data is processed in Section \ref{section:data_analysis}. Next, we have the key results of this paper: we evaluate the effectiveness of the beam model in quantitatively accounting for the mismatch attenuation in Section \ref{section:mismatch}. We conclude in Section \ref{section_conclusion}. 

	\section{Summary of the beam model} \label{section:beam_model}
	We use the beam model of Doppler backscattering in \cite{Hall-Chen:beam_model_DBS:2022} to account for the effect of mismatch, evaluating the applicability of the model to spherical tokamak plasmas. This model uses the reciprocity theorem \cite{Piliya:reciprocity:2002, Gusakov:scattering_slab:2004} in conjunction with beam tracing \cite{Pereverzev:Beam_Tracing:1996, Pereverzev:Beam_tracing:1998}, enabling us to evaluate the backscattered signal in general geometry.
	
	To introduce notation used in this paper, we briefly summarise the model over the rest of this section. We linearise the dielectric tensor, such that it has a large part associated with the equilibrium $\bm{\epsilon}_{eq}$ and a small part related to the turbulent density fluctuations $\bm{\epsilon}_{tb}$, 
	\begin{equation} \label{eq:density_tb}
		\bm{\epsilon}_{tb} = \frac{\delta n_e}{n_e} (\bm{\epsilon}_{eq} -\bm{1}).
	\end{equation}	
	Here $\bm{1}$ is the $3 \times 3$ identity matrix, while $n_e$ and $\delta n_e$ are equilibrium and fluctuating parts of the electron density, respectively. The two pieces of the dielectric tensor, $\bm{\epsilon}_{eq}$ and $\bm{\epsilon}_{tb}$, determine two distinct pieces of the microwave electric field: the beam electric field $\mathbf{E}_{b}$,
	\begin{equation} \label{eq:Maxwell_eq}
		\frac{c^2}{\Omega^2} \nabla \times (\nabla \times \mathbf{E}_{b}) = \bm{\epsilon}_{eq} \cdot \mathbf{E}_{b},
	\end{equation}
	and the scattered electric field $\mathbf{E}_{s}$,
	\begin{equation} \label{eq:Maxwell_tb}
		\frac{c^2}{\Omega^2} \nabla \times (\nabla \times \mathbf{E}_{s}) 
		= \bm{\epsilon}_{tb} \cdot \mathbf{E}_{b} 
		+ \bm{\epsilon}_{eq} \cdot \mathbf{E}_{s},
	\end{equation}	
	where $c$ is the speed of light and $\Omega$ is the angular frequency of the probe beam. We use various diagnostics, such as Thomson scattering and Mirnov coils, to find $\bm{\epsilon}_{eq}$. With the equilibrium dielectric tensor, we use beam tracing to find the probe beam's electric field $\mathbf{E}_{b}$ and the reciprocity theorem \cite{Piliya:reciprocity:2002, Gusakov:scattering_slab:2004} to determine the relevant scattered electric field $\mathbf{E}_s$. 
	
	Beam tracing for fusion plasmas was developed by Pereverzev, Poli, and collaborators \cite{Pereverzev:Beam_tracing:1998, Poli:Torbeam:2001}. In this work, we use the formalism presented in Appendix A of our earlier work \cite{Hall-Chen:beam_model_DBS:2022}, which is equivalent to, but an alternative of, Pereverzev's derivation \cite{Pereverzev:Beam_tracing:1998}. Beam tracing evolves a Gaussian beam by finding its envelope around a central ray $\mathbf{q} (\tau)$, where $\tau$ is a scalar coordinate that determines the position along the ray. The probe beam's electric field is given by
	\begin{equation} \label{eq:beam_field_final}
	\eqalign{
		\mathbf{E}_{b}
		&=
		A_{ant} \hat{\mathbf{e}} \exp(\rmi \phi_G + \rmi \phi_P) \left[ \frac{\det (\textrm{Im} [\bm{\Psi}_w])  }{\det (\textrm{Im} [\bm{\Psi}_{w,ant} ])} \right]^{\frac{1}{4}} \\
		&\times
		\sqrt{\frac{g_{{ant}}}{g}} \exp \left(\rmi s + \rmi \mathbf{K}_w \cdot \mathbf{w} + \frac{\rmi}{2} \mathbf{w} \cdot \bm{\Psi}_w \cdot \mathbf{w}  \right)
		.
	}
	\end{equation}	
	Here $A_{ant}$ is the amplitude of the probe's electric field at the antenna, $\phi_G$ is the Gouy phase, $\phi_P$ is the phase associated with the changing polarisation when propagating through a plasma, $\hat{\mathbf{e}}$ is the polarisation of the beam, $\mathbf{g} = \rmd \mathbf{q} / \rmd \tau$ is the group velocity, $\mathbf{w}$ is the position vector perpendicular to the central ray, $s$ is the eikonal, the complex symmetric matrix $\bm{\Psi}_w$ contains the beam width and curvature, and $\mathbf{K}$ is the wavevector of the beam. The subscript $_{ant}$ denotes that the quantity is evaluated at the antenna, while $_{w}$ indicates that the quantity is projected in the direction perpendicular to the central ray of the beam, that is, in the $\bm{1} - \hat{\mathbf{g}} \hat{\mathbf{g}}$ direction, where $\hat{\mathbf{g}} = \mathbf{g} / g$. The amplitude of the electric field's dependence on $g ^ {- 1 / 2}$ and $\left[ \det (\textrm{Im} [\bm{\Psi}_w]) \right]^{1/4}$ ensures conservation of energy. The eikonal is given by
	\begin{equation} \label{eq:eikonal_s}
		s = \int_0^\tau K_g (\tau') g (\tau') \ \rmd \tau' ,
	\end{equation}
	with $K_g (\tau) = \mathbf{K} (\tau) \cdot \hat{\mathbf{g}} (\tau)$ being the projection of the wavevector along the ray. In the orthonormal basis ($\hat{\mathbf{x}}$, $\hat{\mathbf{y}}$, $\hat{\mathbf{g}}$), the vectors $\mathbf{w}$ and $\mathbf{K}_w$, as well as the matrix $\bm{\Psi}_w$, are given by 
	\begin{eqnarray}
		\mathbf{w} =
		\left( \begin{array}{c}
			w_{x} \\
			w_{y} \\
			0   \\
		\end{array} \right) ,
	\end{eqnarray}
	\begin{eqnarray}
		\mathbf{K}_w =
		\left( \begin{array}{c}
			K_{x} \\
			K_{y} \\
			0   \\
		\end{array} \right) ,
	\end{eqnarray}
	\begin{eqnarray}
		\bm{\Psi}_w =
		\left( \begin{array}{ccc}
			\Psi_{xx} & \Psi_{xy} & 0 \\
			\Psi_{yx} & \Psi_{yy} & 0 \\
			0		  & 0		 & 0 \\
		\end{array} \right) ,
	\end{eqnarray}
	where we define $\det (\bm{\Psi}_w) = \Psi_{xx} \Psi_{yy} - \Psi_{xy}^2$. The real part of $\bm{\Psi}_w$ is responsible for the curvature of the Gaussian beam, while its imaginary part gives the characteristic decay width of the Gaussian envelope. In general, the real and imaginary parts are not simultaneously diagonalisable. When separately diagonalised, the eigenvalues of the real part are
	\begin{equation} \label{eq:Psi_real}
		\left[ \textrm{Re} \left( \bm{\Psi}_{w}  \right) \right]_{\alpha \alpha} 
		= \frac{K^3}{K_g^2} \frac{1}{R_{b,\alpha}}
		,
	\end{equation}  
	where $R_{b,\alpha}$ are the principal radii of curvature of the beam front, while the eigenvalues of the imaginary part are
	\begin{equation} \label{eq:Psi_imag}
		\left[ \textrm{Im} \left( \bm{\Psi}_{w}  \right) \right]_{\alpha \alpha}
		= \frac{2}{W_{\alpha}^2} ,
	\end{equation}	
	where $W_{\alpha}$ are the principal beam widths. We solve for $\mathbf{E}_b$, given in equation (\ref{eq:beam_field_final}), with our open-source beam-tracing code, Scotty \cite{Hall-Chen:beam_model_DBS:2022, Hall-Chen:Scotty:2022}.
	
\begin{figure}
		\centering
		\begin{tikzpicture}[>=latex]
		\draw[style=help lines] (0,0) (3,2);
		
		\coordinate (vec1) at (90:2.5);
		\coordinate (vec2) at (0:2.5);
		\coordinate (vec3) at (200:2.5);
		\coordinate (vec4) at (110:2.5);
		\coordinate (vec5) at (140:2.5);
		\coordinate (vec6) at (180:2.5);
		
		\draw[->,thick,black] (0,0) -- (vec1) node[left] {$\hat{\mathbf{g}}$};
		\draw[->,thick,black] (0,0) -- (vec2) node[right] {$\hat{\mathbf{x}}$};
		\draw[->,thick,blue] (0,0) -- (vec3) node [below] {$\hat{\mathbf{b}}$};
		\draw[->,thick,blue] (0,0) -- (vec4) node [left] {$\hat{\mathbf{u}}_{1}$};
		\draw[->,thick,red] (0,0) -- (vec5) node [left] {$\mathbf{K}$};
		\draw[dashed,thick,black] (0,0) -- (vec6) ;
		
		\draw [->,black,thick,domain=90:110] plot ({1.5*cos(\x)}, {1.5*sin(\x)}) node[above right] {$\theta$};
		\draw [->,black,thick,domain=110:140] plot ({1.0*cos(\x)}, {1.0*sin(\x)});
		\node[above] at (-0.75,0.75) {$\theta_m$};
		\draw [->,black,thick,domain=180:200] plot ({1.5*cos(\x)}, {1.5*sin(\x)});
		\node[below] at (-1.65,0.0) {$\theta$};
		
		\draw (0,.3)-|(.3,0);
		\draw ({0.3*cos(200)}, {0.3*sin(200)}) -- ({0.3*cos(200)-0.3*cos(70)}, {0.3*sin(200)+0.3*sin(70)}) -- ({0.3*cos(110)}, {0.3*sin(110)});
		
		\draw[blue,thick] (0,-0.5) circle (0.2cm) node[below right] {$\hat{\mathbf{u}}_{2} = \hat{\mathbf{y}}$};
		\draw[blue,thick] (-0.14,-0.64) -- (0.14,-0.36);
		\draw[blue,thick] (0.14,-0.64) -- (-0.14,-0.36);
		
		\end{tikzpicture} \\
		
		\caption{Bases for $\mathbf{k}_\perp$ and $\mathbf{w}$. Here $\hat{\mathbf{b}}$ and $\hat{\mathbf{g}}$ are unit vectors parallel to the magnetic field and group velocity, respectively, $\mathbf{K}$ is the probe beam's wavevector, and $\theta_m$ is the mismatch angle. The beam-aligned coordinates are in the basis $\left( \hat{\mathbf{g}}, \hat{\mathbf{x}}, \hat{\mathbf{y}} \right)$ while that of the field-aligned coordinates are $\left( \hat{\mathbf{b}}, \hat{\mathbf{u}}_1, \hat{\mathbf{u}}_2 \right)$; the angle between these two bases is related to $\theta$.} 
		\label{fig:basis}
	\end{figure}
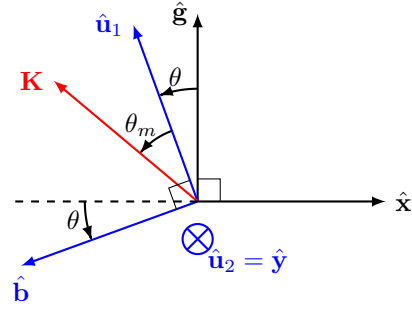

	Once we determine the probe beam's electric field, we find the backscattered amplitude via reciprocity \cite{Piliya:reciprocity:2002, Gusakov:scattering_slab:2004},
	\begin{equation} \label{eq:reciprocity_theorem}
		A_r 
		=\frac{\Omega \rmi}{ 2 c } \int \mathbf{E}^{(+)} \cdot \bm{\epsilon}_{tb} \cdot \mathbf{E}_{b} \ \rmd V .
	\end{equation}
	For cold plasmas, it turns out that the reciprocal electric field $\mathbf{E}^{(+)}$ has almost the same form as $\mathbf{E}_b$ \cite{Hall-Chen:beam_model_DBS:2022}. To evaluate the integrals, we need to take into account the fact that turbulence structures are elongated along the magnetic field. For this reason, we define an orthonormal basis in the magnetic field's frame, which helps us to express the components of the density fluctuations. We have $(\hat{\mathbf{u}}_1, \hat{\mathbf{u}}_2, \hat{\mathbf{b}})$ at each point along the beam. Here $\hat{\mathbf{b}}$ and $\hat{\mathbf{g}}$ are the unit vectors along the external magnetic field $\mathbf{B}$ and group velocity $\mathbf{g}$, respectively; while $\hat{\mathbf{u}}_1$ and $\hat{\mathbf{u}}_2$ are given by
	\begin{equation} \label{eq:kperp1}
		\hat{\mathbf{u}}_{1} 
		= \frac{ (\hat{\mathbf{b}} \times \hat{\mathbf{g}}) \times \hat{\mathbf{b}} }{ |(\hat{\mathbf{b}} \times \hat{\mathbf{g}}) \times \hat{\mathbf{b}}| },
	\end{equation}
	and
	\begin{equation} \label{eq:kperp2}
		\hat{\mathbf{u}}_{2} 
		= \frac{ \hat{\mathbf{b}} \times \hat{\mathbf{g}} }{ | \hat{\mathbf{b}} \times \hat{\mathbf{g}} | } .
	\end{equation}	
	We also associate another orthonormal basis, that of the beam's reference frame $(\hat{\mathbf{x}}, \hat{\mathbf{y}}, \hat{\mathbf{g}})$, with each point along the beam to express the beam characteristics. We align the $(\hat{\mathbf{x}}, \hat{\mathbf{y}}, \hat{\mathbf{g}})$ basis with the $(\hat{\mathbf{u}}_1, \hat{\mathbf{u}}_2, \hat{\mathbf{b}})$ basis as shown in Figure \ref{fig:basis}. We choose $\hat{\mathbf{y}} = \hat{\mathbf{u}}_{2}$ and denote projection in that direction with the subscript $_y$. The other basis which is perpendicular to both $\hat{\mathbf{g}}$ and $\hat{\mathbf{y}}$ is
	\begin{equation}
		\hat{\mathbf{x}}
		= \frac{ \hat{\mathbf{y}} \times \hat{\mathbf{g}} }{ |\hat{\mathbf{y}} \times \hat{\mathbf{g}} | }
		= \frac{ \hat{\mathbf{u}}_{2} \times \hat{\mathbf{g}} }{ |\hat{\mathbf{u}}_{2} \times \hat{\mathbf{g}} | }.
	\end{equation}
	The detailed rationale for this choice of coordinate systems is given in our previous work \cite{Hall-Chen:beam_model_DBS:2022}.
	
	To make analytical progress, we need to make a few assumptions about the mismatch angle. We consider two situations, the \emph{small-mismatch} and \emph{large-mismatch} orderings. These names refer to two different orderings typical of but not exclusive to conventional and spherical tokamaks, respectively. In the \emph{small-mismatch} ordering, the mismatch angle is small throughout the path of the beam, $\theta_m \sim W / L$, where $W$ is the characteristic beam width and $L$ is the inhomogeneity length scale. Hence, the signal is largely localised to the cut-off location. This backscattered signal from the cut-off is attenuated by mismatch, which we determine with our model. In the \emph{large-mismatch} ordering, the mismatch angle is allowed to be up to order unity along the path of the beam, but on at least one point along the beam, it must be zero. The signal is then localised to the points of zero mismatch, which may not be at the cut-off. We showed the result obtained from the \emph{small-mismatch} ordering are still applicable even in situations which are moderately in the \emph{large-mismatch} regime \cite{Hall-Chen:beam_model_DBS:2022}. 
	
	Consequently, we now consider the small mismatch angle limit. After several integrals across and along the beam, and in one of the directions of $\mathbf{k}_\perp$, we express the integral along the other direction of $\mathbf{k}_\perp$ as an integral along the arc length $l$ of the central ray and equation (\ref{eq:reciprocity_theorem}) thus gives
	\begin{equation} \label{eq:A_r2_final_cleaned}
		\frac{p_r}{P_{ant}} 
		=
		\frac{\sqrt{\pi^3} e^4}{2 c^2 \Omega^2 \epsilon_{0}^2 m_e^2 \bar{W}_{y} }
		\int 
		F_t	
		\left<
			\delta n_{e}^2 (t)
		\right>_t
		\widetilde{C}_l (\omega)
		\ \rmd l .
	\end{equation}
	Here $p_r$ is the backscattered spectral density, $P_{ant}$ is the power that the antenna emits, $\left<\delta n_{e}^2 (t)\right>_t$ is the time-averaged square of the amplitude of electron density fluctuations, $\widetilde{C}_l (\omega)$ is the correlation function, which will be defined later in this section, and $F_t$ is instrumentation weight at every point along the ray. One can also think of $F_t$ as a function that weighs measurements of turbulence at various locations. Other authors have different names for the instrumentation weight and related quantities: spatial resolution \cite{Gusakov:scattering_slab:2004}, instrumentation response function \cite{Lechte:2D_fullwave:2012, Lechte:2D_fullwave:2017}, weighting function \cite{Bulanin:spatial_spectral_resolution:2006}, or filter function \cite{RuizRuiz:RCDR:2022}. We introduce the notation in equation (\ref{eq:A_r2_final_cleaned}) throughout the rest of this section. Here $p_r$ is the spectral density, $l$ is the arc-length along the central ray, $\bm{M}_w$ is the symmetric modified $\bm{\Psi}_w$ matrix, given by
	\begin{eqnarray}
		\bm{M}_w =
		\left(
		\begin{array}{ccc}
			M_{xx} & M_{xy} & 0\\
			M_{xy} & M_{yy} & 0\\
			0 & 0 & 0\\
		\end{array}
		\right) ,
	\end{eqnarray}
	where
	\begin{equation}
		\fl
		\eqalign{
		M_{xx} = \Psi_{xx}
		&- \frac{K_g}{\cos \theta} \bigg(
		\frac{\sin \theta}{g} \frac{\rmd \theta}{\rmd \tau} 
		- \boldsymbol{\kappa} \cdot \hat{\mathbf{x}} \sin \theta \\
		&+ \hat{\mathbf{x}} \cdot \nabla \hat{\mathbf{b}} \cdot \hat{\mathbf{g}}
		- \hat{\mathbf{x}} \cdot \nabla \hat{\mathbf{b}} \cdot \hat{\mathbf{x}} \tan \theta
		\bigg) , }
	\end{equation}
	\begin{equation}
		\fl
		\eqalign{
		M_{xy} = \Psi_{xy}
		&- \frac{K_g}{\cos \theta} \bigg(
		- \boldsymbol{\kappa} \cdot \hat{\mathbf{y}} \sin \theta
		+ \hat{\mathbf{y}} \cdot \nabla \hat{\mathbf{b}} \cdot \hat{\mathbf{g}} \\
		&+ \frac{\sin \theta \tan \theta}{g} \frac{\rmd \hat{\mathbf{x}}}{\rmd \tau} \cdot \hat{\mathbf{y}}
		- \hat{\mathbf{y}} \cdot \nabla \hat{\mathbf{b}} \cdot \hat{\mathbf{x}} \tan \theta
		\bigg)
		, }
	\end{equation}
	and
	\begin{equation}
		M_{yy} = \Psi_{yy}.
	\end{equation}
	Here the subscripts give the relevant directions of $\bm{M}_w$, shown in Figure \ref{fig:basis}; the angle
	\begin{equation}
		\cos \theta = \hat{\mathbf{g}} \cdot \hat{\mathbf{u}}_1
	\end{equation}
	is related to the angle between the group velocity and magnetic field, also shown in Figure \ref{fig:basis}, and
	\begin{equation}
		\boldsymbol{\kappa} = \frac{1}{g} \frac{\rmd \hat{\mathbf{g}}}{\rmd \tau}
	\end{equation}
	is the curvature of the central ray. This ray curvature should not be confused with the wavefront curvature, $R_b$. In general, the real and imaginary parts of $\bm{M}_w$ are not simultaneously diagonalisable, as was the case for $\bm{\Psi}_w$. In this formulation of $\bm{M}_w$, we have used the Bragg condition to evaluate $k_{\perp,1}$ along the ray, and $k_{\perp,2} = 0$; the argument for this was presented in our previous work \cite{Hall-Chen:beam_model_DBS:2022}. To simplify $\bm{M}_w$, we neglect terms that are small in mismatch angle \cite{Hall-Chen:beam_model_DBS:2022}, getting
	\begin{equation}
		M_{xx} \simeq \Psi_{xx}
		- K \left( \hat{\mathbf{b}} \cdot \nabla \hat{\mathbf{b}} \cdot \hat{\mathbf{g}} \right)
		,
	\end{equation}
	\begin{equation}
		M_{xy} \simeq \Psi_{xy}
		- K \frac{\left[ \left( \hat{\mathbf{b}} \times \hat{\mathbf{g}} \right)  \cdot \nabla \hat{\mathbf{b}} \cdot \hat{\mathbf{g}} \right]}{\left| \hat{\mathbf{b}} \times \hat{\mathbf{g}} \right|}
		,
	\end{equation}
	and
	\begin{equation}
		M_{yy} = \Psi_{yy}.
	\end{equation}
	Note that the corrections are related to the curvature of field lines and magnetic shear, not the curvature of the cut-off.		
	
	In equation (\ref{eq:A_r2_final_cleaned}), we use the shorthand
	\begin{equation}
	\eqalign{
		&\widetilde{C}_l (\omega) = 
			\widetilde{C} \Big(
			\mathbf{r}=\mathbf{q} (\tau (l)), k_{\perp,1} = - 2 K (\tau (l)), \\
		&\qquad \qquad
			k_{\perp,2} = 0,  
		\Delta u_{\parallel} = 0, 
			\omega
		\Big), } \\
	\end{equation}	
	where $u_{\parallel}$ is the arc length along magnetic field lines and $\widetilde{C}$ is the Fourier-transformed correlation function, given by
	\begin{equation}
	\fl
	\eqalign{
		& \widetilde{C} (\mathbf{r}, k_{\perp,1}, k_{\perp,2}, \Delta u_{\parallel}, \omega) \\
		& \qquad = \left<
			\delta n_e^2 (\mathbf{r}, t)
		\right>_t^{-1} \\ 
		& \qquad \times \int
		\delta \tilde{n}_e \left( k_{\perp, 1}, k_{\perp, 2}, u_{\parallel} + \Delta u_\parallel, \omega \right) \\
		& \qquad \times \delta \tilde{n}_e^* \left( k_{\perp, 1}', k_{\perp, 2}', u_{\parallel}, \omega \right) \\
		& \qquad \times \exp \left[
		\rmi (k_{\perp, 1} - k_{\perp, 1}') u_1
		\right] \\
		& \qquad \times \exp \left[
		\rmi (k_{\perp, 2} - k_{\perp, 2}') u_2
		\right]
		\ \rmd k_{\perp, 1}' \ \rmd k_{\perp, 2}' .
	}
	\end{equation}
	We have used the Bragg condition to get $k_{\perp,1} = - 2 K (\tau (l))$; in earlier work \cite{Hall-Chen:beam_model_DBS:2022}, we had
	\begin{equation} \label{eq:Bragg_condition_full}
		k_{\perp, 1} \cos \theta (\tau) = - 2 K_g (\tau) .
	\end{equation}	
	Neglecting terms that are small in mismatch, we get
	\begin{equation} \label{eq:Bragg_condition}
		k_{\perp, 1} \simeq - 2 K (\tau) ,
	\end{equation}	
	which is how the Bragg condition is typically presented in the literature: at every point along the ray, there is a specific $k_{\perp,1}$ that is responsible for backscattering, and its value is determined solely by the wavenumber at that point. Moving forward, we define the inverse of $\bm{M}_w$ as
	\begin{eqnarray}
	\bm{M}_w^{-1}
	&=
	\left( \begin{array}{ccc}
	M_{xx}^{-1} & M_{xy}^{-1} & 0 \\
	M_{yx}^{-1} & M_{yy}^{-1} & 0 \\
	0		  & 0		 & 0 \\
	\end{array} \right) \\
	&=
	\left(
	\begin{array}{cc}
	&
	\left(
	\begin{array}{cc}
	M_{xx} & M_{xy} \\
	M_{yx} & M_{yy}
	\end{array}
	\right) ^{-1}
	
	\begin{array}{c}
	0 \\
	0 \\
	\end{array} \\
	
	& \begin{array}{cc}
	\ \ \ \ 0 \ \ \ \ & 0~~
	\end{array}

	\begin{array}{c}
	\ \ \ \ \ \ \ 0
	\end{array}
	
	\end{array} \right) .
	\end{eqnarray}
	It is important to bear in mind that $M_{ij}^{-1}$ is the $i j$ component of $\bm{M}_w^{-1}$, and not $1 / M_{ij}$.	
	
	We split the integrand in equation (\ref{eq:A_r2_final_cleaned}) into two parts. The first is the turbulence we are trying to measure, given by
	\begin{equation}
		\left<
			\delta n_{e}^2 (t)
		\right>_t
		\widetilde{C}_l (\omega).
	\end{equation}
	The turbulence spectrum may contribute to the instrumentation weight. For electrostatic turbulence, references \cite{Schekochihin:spectrum:2008, Schekochihin:spectrum:2009, Barnes:spectrum:2011} suggest that the spectrum is of the form $\widetilde{C} (\mathbf{r}, k_{\perp,1}, k_{\mu,2}, \Delta u_\parallel, \omega) \propto k_\perp^{- 10 / 3}$ or $\widetilde{C} (\mathbf{r}, k_{\perp,1}, k_{\mu,2}, \Delta u_\parallel, \omega) \propto k_\perp^{- 13 / 3}$, depending on the size of $k_{\perp} \rho_i$. Since the backscattered $k_{\perp,1}$ is related to the beam's wavevector via the Bragg condition, equation (\ref{eq:Bragg_condition}), the turbulence spectrum localises signal to the cut-off, which we nominally take to be the point along the ray where the beam wavenumber $K$ is at a minimum. While the exact turbulence spectrum differs from experiment to experiment, this power law enables us to qualitatively study the effect of turbulence on measurements of mismatch attenuation. The second part of equation (\ref{eq:A_r2_final_cleaned}) is the instrumentation weight, which we split into various factors
	\begin{equation} \label{eq:localisation}
		F_t = F_p  F_r F_b F_m .		
	\end{equation}
	Here we have the polarisation factor, which we referred to as $\varepsilon$ in previous work \cite{Hall-Chen:beam_model_DBS:2022},
	\begin{equation}
		F_p = \varepsilon = 
		\frac{\Omega^4 \epsilon_{0}^2 m_e^2}{e^4 n_{e,\mu}^2}
		|\hat{\mathbf{e}}^* \cdot (\bm{\epsilon}_{eq} -\bm{1}) \cdot \hat{\mathbf{e}}|^2 ,
	\end{equation}
	ray piece,
	\begin{equation}\label{eq:ray_factor}
		F_r = 
		\frac{ g_{ant}^{2} }{ g^{2} } ,
	\end{equation}	
	beam piece,
	\begin{equation}\label{eq:beam_factor}
	F_b = 
	\frac{
		\bar{W}_{y}
		\det \left[
		\textrm{Im} \left( \bm{\Psi}_{w} \right) 
		\right] 
	}{
		\sqrt{2}
		\left| 
		\det \left[ \bm{M}_{w} \right] 
		\right| 
		\left[
		- \textrm{Im} \left( M_{yy}^{-1} \right)
		\right]^{\frac{1}{2}}  
	} ,
	\end{equation}
	and mismatch piece,
	\begin{equation}\label{eq:mismatch_attenuation}
	F_m = 
	\exp\left[ 
		- 2 \frac{
		\theta_m^2 (l)			
		}{
		\left[
			\Delta \theta_m (l)
		\right]^2
		}
		\right] ,
	\end{equation}
	where $\Delta \theta_m (l)$ is the characteristic width of the mismatch attenuation, given by
	\begin{equation} \label{eq:delta_theta_m}
		\Delta \theta_{m}
		= 
		\frac{ 1 }{ K } 
		\left(
			\frac{
				\textrm{Im} \left( M_{yy}^{-1} \right)
			}{
			\left[ \textrm{Im} \left( M_{xy}^{-1} \right) \right]^2
			-
			\textrm{Im} \left( M_{xx}^{-1} \right) \textrm{Im} \left( M_{yy}^{-1} \right)
			}
		\right)^{\frac{1}{2}} ,
	\end{equation} 	
	The physical intuition relating to the various contributions to $\Delta \theta_m$ is given in our earlier work \cite{Hall-Chen:beam_model_DBS:2022}. Having introduced the beam model and its associated notation in this section, we describe our experimental approach in the next.

	\section{Analysis of experimental data} \label{section:data_analysis}
	In this section, we explain what data was studied and how we analysed it. This begins in subsection \ref{subsection:shot_parameters} with an introduction to the various MAST and MAST-U shots that were used for this paper. We then describe in subsection \ref{subsection:equilibrium} what equilibrium data was needed. The geometry of the MAST DBS is detailed in subsection \ref{subsection:DBS_geometry}, while that of MAST-U is given in previous work \cite{Rhodes:DBS:2022}. Subsequently, we state and explain the initial beam parameters used for our beam tracing simulations of the MAST DBS in \ref{appendix:beam_parameters}. Finally, we detail how we analysed experimental DBS data in subsection \ref{subsection:DBS_analysis}.
	
	\subsection{Shot parameters} \label{subsection:shot_parameters}
	To study the effect of mismatch attenuation, we varied the toroidal launch angle and thus the mismatch angle. Pulses were repeated with the same plasma current, shape, and heating waveforms, such that the equilibria were as similar as experimentally possible. The toroidal angle was varied shot to shot. We studied two sets of repeated shots on MAST (shot group 1) and MAST-U (shot group 2), respectively. Details of these shots are given in Table \ref{table:shotgroup}. Shots in group 1 were neutral-beam-injection driven with plasma currents $\sim 800$ kA and shots in group 2 were ohmic with plasma currents $\sim 70$ kA. 
	%

\begin{figure*}
		\centering
		\includegraphics[width=0.75\linewidth]{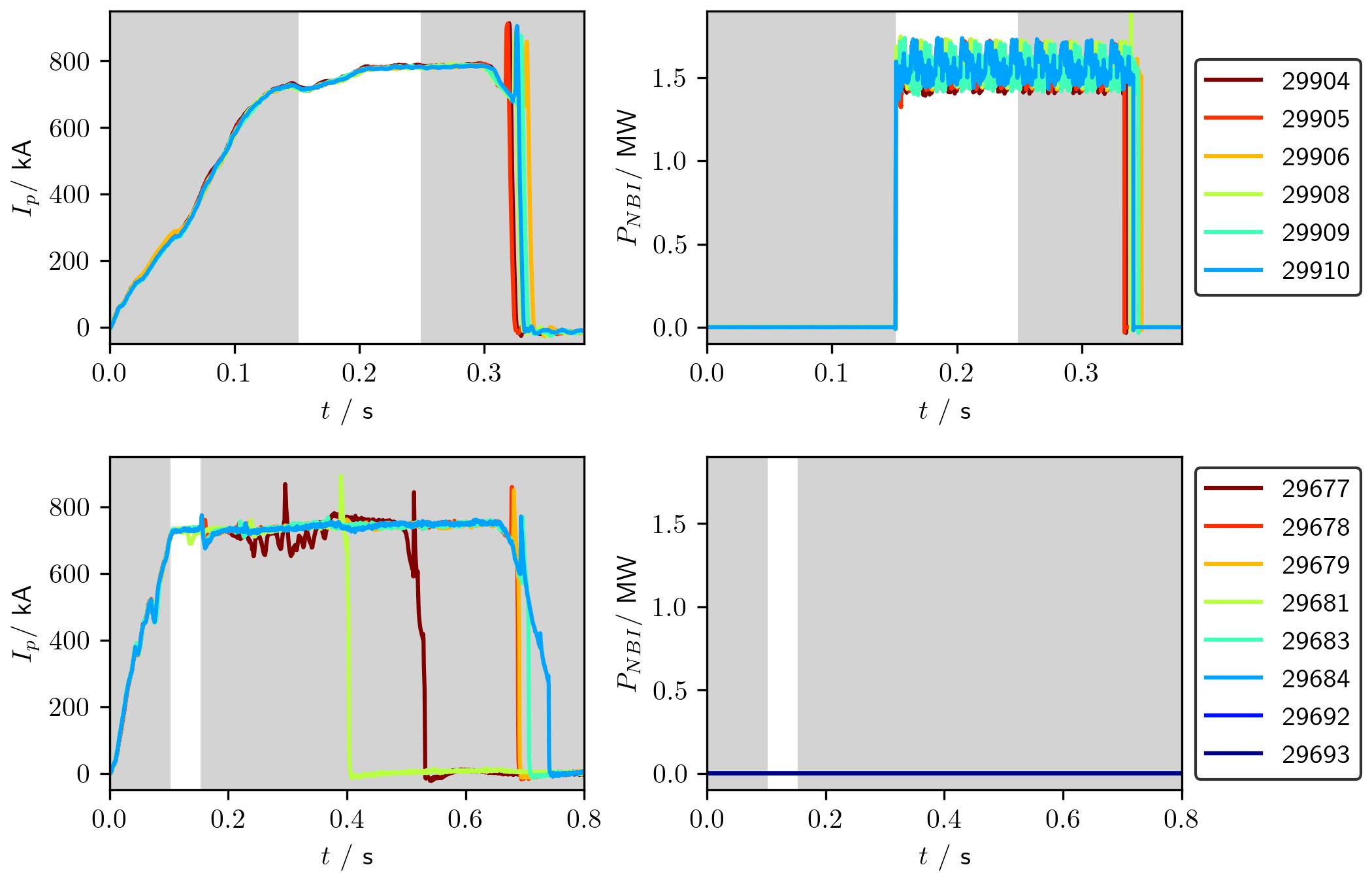}
		\caption{Shot group 1 (top) and shot group 2 (bottom). Plasma current (left) and total NBI power (right); shot group 2 was Ohmic and had no neutral beams. We studied data from the time intervals shaded in white, ignoring data from times shaded in grey. Plasma current and neutral beam power from shot group 1 were previously published \cite{Hillesheim:DBS_MAST:2015}.}
		\label{fig:Ip_and_NBI}
	\end{figure*}

\begin{figure*}
		\includegraphics[width=0.99\textwidth]{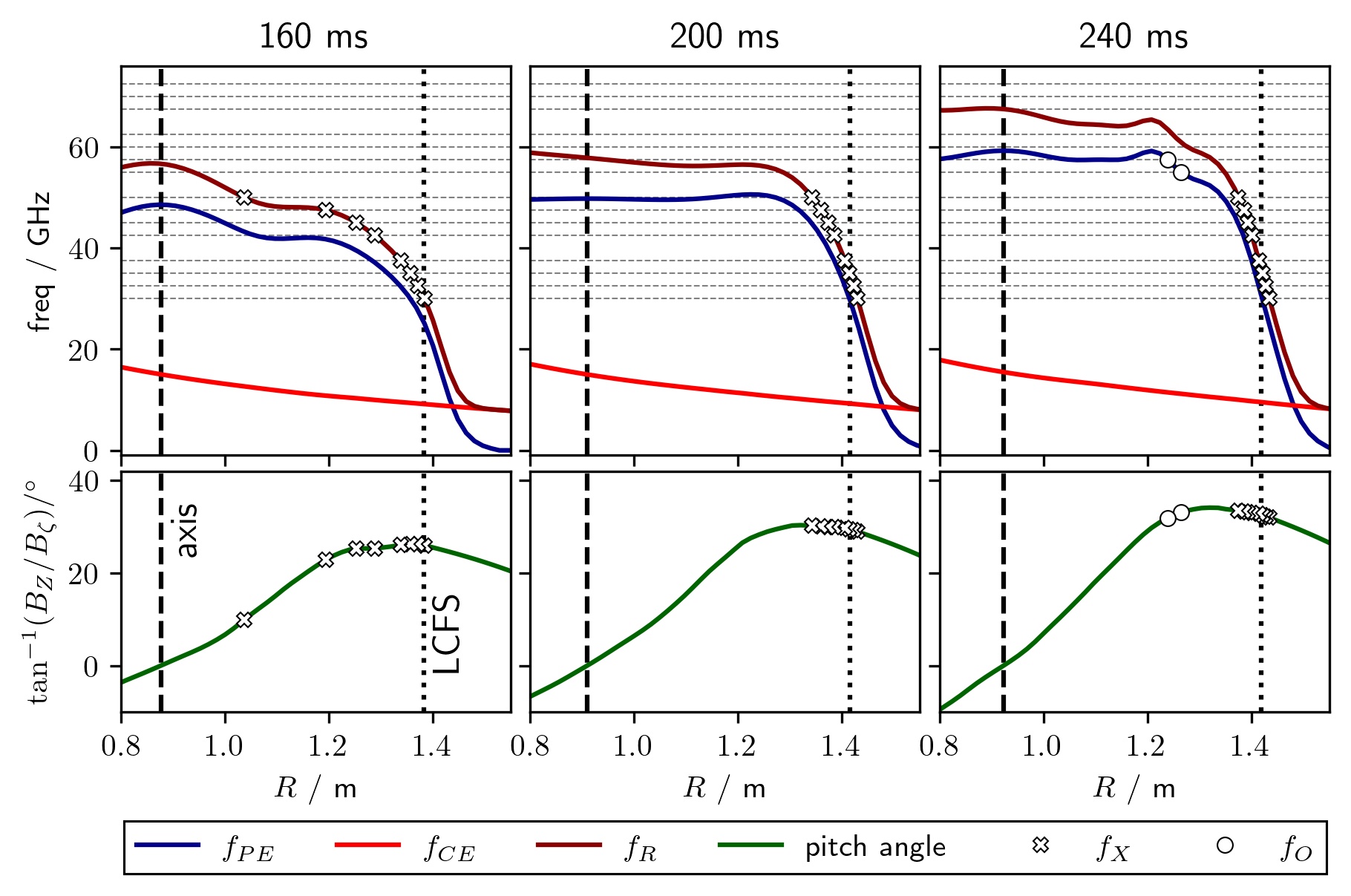}
		\caption{Properties of the plasma equilibrium for MAST shot group 1. MAST DBS frequencies indicated by the horizontal dashed lines and points (crosses for X-mode and circles for O-mode). Here $f_{PE}$ is the plasma frequency, $f_{CE}$ is the cyclotron frequency, $f_{R}$ is the X-mode cut-off frequency, $\tan^{-1} (B_Z / B_\zeta)$ is the pitch angle, $B_Z$ is a poloidal component of the magnetic field, and $B_\zeta$ the toroidal component. The values of these quantities on the midplane are plotted as a function of radius, $R$. }
		\label{fig:equilibria}
	\end{figure*}

	\begin{table} 
		\centering
		
		\begin{tabular}{ |p{3cm}||p{1cm}|p{1cm}|  }
			\hline
			\multicolumn{3}{|c|}{Shot group 1 (MAST)} \\
			\hline
			Shot number & $\varphi_{rot}$ & $\varphi_{tilt}$ \\
			\hline
			29904 & $-4^{\circ}$ & $-4^{\circ}$  \\
			29905 & $-5^{\circ}$ & $-4^{\circ}$  \\
			29906 & $-6^{\circ}$ & $-4^{\circ}$  \\
			29908 & $-3^{\circ}$ & $-4^{\circ}$  \\
			29909 & $-2^{\circ}$ & $-4^{\circ}$  \\
			29910 & $-1^{\circ}$ & $-4^{\circ}$  \\ [1ex] 
			\hline
			\hline
			\multicolumn{3}{|c|}{Shot group 2 (MAST-U)} \\
			\hline
			Shot number & $\varphi_{tor}$ & $\varphi_{pol}$ \\
			\hline
			45288 & $2.2^{\circ}$ & $-2.5^{\circ}$   \\
			45289 & $0.9^{\circ}$ & $-2.5^{\circ}$   \\
			45290 & $3.5^{\circ}$ & $-2.5^{\circ}$   \\
			45291 & $4.6^{\circ}$ & $-2.5^{\circ}$   \\
			45292 & $4.6^{\circ}$ & $-2.5^{\circ}$   \\															
			45293 & $0.1^{\circ}$ & $-2.5^{\circ}$   \\ [1ex] 		
			\hline
		\end{tabular}
		
		\caption{Summary of shots examined in this paper, together with the DBS mirror angles (MAST) and launch angles (MAST-U). For shot group 1, the Q-band (30--50 GHz) was in X-mode and the V-band (55--75 GHz) was in O-mode. Shot group 2 only had Q-band measurements in X-mode. Here $\varphi_{rot}$ and $\varphi_{tilt}$ are the angles of the steering mirror, see subsection \ref{subsection:DBS_geometry}. Here $\varphi_{pol}$ and $\varphi_{tor}$ are the poloidal and toroidal launch angles, respectively, defined in equation (\ref{eq:launch_K}). }
		\label{table:shotgroup}
	\end{table}	
	In shot group 1, the shots were repeated with the same conditions from start to end, Figure \ref{fig:Ip_and_NBI}. The shots in group 2 were also repeated shots, with significant differences only after 150 ms. In shot group 1, the MAST DBS was configured such that the V-band launched O-mode polarisation and the Q-band launched X-mode polarisation. In shot group 2, the Q-band launched X-mode polarised beams.
	As density and plasma current increased as the shots in group 1 progressed, the Q-band DBS on MAST moved from measuring deep into the core to the edge, Figure \ref{fig:equilibria}. This enables us to compare the calculated mismatch attenuation with data from across multiple different plasma equilibria with the same DBS configuration, as discussed in Subsection \ref{subsection:data_Q}.
		


	
	\subsection{Magnetic equilibria and density profiles} \label{subsection:equilibrium}
	In order to determine the equilibrium dielectric tensor $\bm{\epsilon}_{eq}$, we need to know the electron density $n_e (\mathbf{r})$ and magnetic field $\mathbf{B} (\mathbf{r})$, both as functions of position. This is made simpler by assuming toroidal symmetry. Hence, we only need to know these quantities as expressed in the poloidal plane, as a function of $R$ and $Z$. We use EFIT \cite{Lao:EFIT:1985, Appel:EFIT:2006}, constrained by the motional Stark effect diagnostic \cite{Conway:MAST_MSE:2010} when possible, to determine the normalised poloidal flux and the toroidal magnetic field $\mathbf{B}_\zeta$. MSE-constrained EFIT was available for shot group 1 (see Table \ref{table:shotgroup}) but not for shot group 2 since there were no neutral beams. Nonetheless, edge-magnetics-only EFIT should be sufficient for shot group 2 for two reasons. First, at the times studied, the plasmas were in L-mode. Secondly, we only evaluated DBS data from scattering near the edge for these shots. In addition to EFIT, we used the Thomson scattering (TS) diagnostic \cite{Scannell:ThomsonScattering:2010}, which gives density $n_e$ at points along its path. The TS electron density data, together with the poloidal flux from EFIT at points along the TS laser beam, we express electron density as a function of flux label $n_e (\psi_p)$. In this work, we used a smoothing spline for interpolation of the density profiles. 
	
	We use the equilibrium dielectric tensor, as determined from EFIT and Thomson scattering, as input for our beam tracing code, Scotty \cite{Hall-Chen:beam_model_DBS:2022}. Scotty first determines the probe beam's electric field via beam tracing. It subsequently calculates the mismatch attenuation and other instrumentation weights. Such calculation only requires the beam parameters and some post-processing with our model, so one is not limited to using Scotty; any beam tracing code, such as Torbeam \cite{Poli:Torbeam:2001}, would suffice.


	\subsection{Geometry of DBS hardware} \label{subsection:DBS_geometry}
	The current DBS system on MAST-U is described in another paper \cite{Rhodes:DBS:2022}. While the previous MAST DBS was also covered in previous work \cite{Hillesheim:DBS_MAST:2015}, we proceed to show its setup in further detail. Detailed system geometry helps set the groundwork for a later section, where we argue that the MAST Q-band has a small systematic error. This manifests as an apparent error in the mirror rotation angle; however, since the same mirror is used for both bands and the V-band data does not have this error, the error must arise from the misalignment of another component in the Q-band quasioptical system. Direct measurement of this error is no longer possible because this DBS system has already been dismantled. 

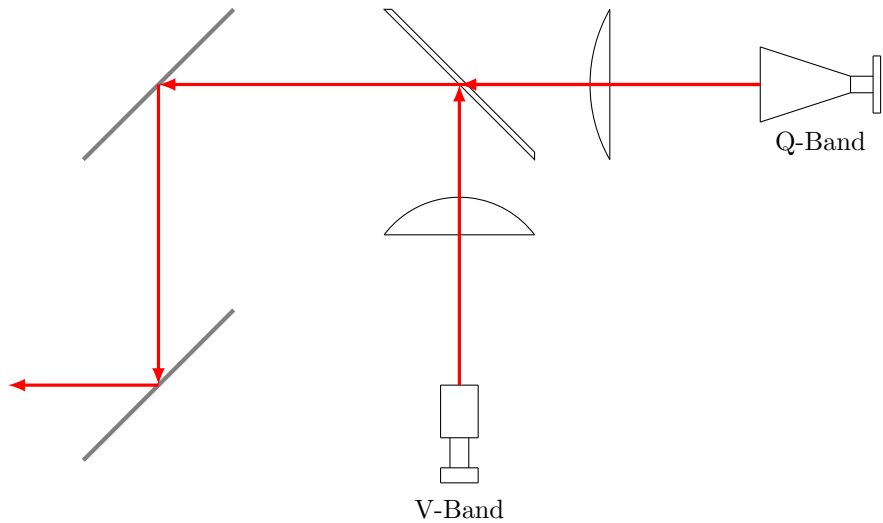
\begin{figure*}
	\centering
	
	\begin{tikzpicture}[>=latex]
	
	\draw[-, thin, black] (4, -0.5) -- (4, 0.5);
	\draw[-, thin, black] (5.2, -0.11) -- (5.2, 0.11);
	\draw[-, thin, black] (5.5, -0.38) -- (5.5, 0.38);	
	\draw[-, thin, black] (5.6, -0.38) -- (5.6, 0.38);	
	\draw[-, thin, black] (4, 0.5) -- (5.2, 0.11)  -- (5.5, 0.11);
	\draw[-, thin, black] (5.5, 0.38) -- (5.6, 0.38);	
	\draw[-, thin, black] (4, -0.5) -- (5.2, -0.11)  -- (5.5, -0.11);
	\draw[-, thin, black] (5.5, -0.38) -- (5.6, -0.38);	
	\node[below=0.2cm, align=flush center] at (4.8,-0.3) {Q-Band};				
				
	\draw[-, thin, black] (-0.25, -4) -- (0.25, -4);
	\draw[-, thin, black] (-0.25, -4.7) -- (0.25, -4.7);
	\draw[-, thin, black] (-0.25, -5.1) -- (0.25, -5.1);	
	\draw[-, thin, black] (-0.25, -5.3) -- (0.25, -5.3);
	\draw[-, thin, black] (-0.25, -4) -- (-0.25, -4.7);	
	\draw[-, thin, black] (-0.125, -4.7) -- (-0.125, -5.1);	
	\draw[-, thin, black] (-0.25, -5.1) -- (-0.25, -5.3);	
	\draw[-, thin, black] (0.25, -4) -- (0.25, -4.7);	
	\draw[-, thin, black] (0.125, -4.7) -- (0.125, -5.1);	
	\draw[-, thin, black] (0.25, -5.1) -- (0.25, -5.3);		
	\node[below=0.1cm, align=flush center] at (0,-5.3) {V-Band};				
	
	\draw[-, thin, black] (-1, -2) -- (1, -2) ;	
	\draw[-, thin, black] (-1,-2) arc(143.13:36.87:1.25);
	
	\draw[-, thin, black] (2, 1) -- (2, -1);	
	\draw[-, thin, black] (2,-1) arc(209.5:150.5:2.031);
			
	\draw[-, thin, black] (-1, 1) -- (-0.9, 1) -- (1,-0.9) -- (1,-1) -- cycle;	
		
	\draw[-, ultra thick, black!50]  (-3,1) -- (-5,-1);
	\draw[-, ultra thick, black!50]  (-3,-3) -- (-5,-5);
		
	\draw[->, very thick, red] (4, 0) -- (0, 0);
	\draw[->, very thick, red] (0, -4) -- (0, 0);
	
	\draw[->, very thick, red] (4, 0) -- (-4, 0);
	\draw[->, very thick, red] (-4, 0) -- (-4, -4);
	\draw[->, very thick, red] (-4, -4) -- (-6, -4);	
	\end{tikzpicture}

	\caption{Previous DBS system on MAST. Simplified 2D schematic of the horns, lenses, adjustable polariser, and mirrors. The Q-band beam is transmitted through the polariser, while the V-band beam is reflected off it; this is why the polarisation of the two bands are always orthogonal to each other. The polariser is adjusted such that one band is in O-mode and the other X-mode. The last mirror here is the steering mirror, which appears in Figure \ref{fig:launch_geometry}. For a 3D view of the experimental layout, refer to Figure 6 of previous work on the MAST DBS \cite{Hillesheim:DBS_MAST:2015}.}
	\label{fig:horn_geometry}
	\end{figure*}

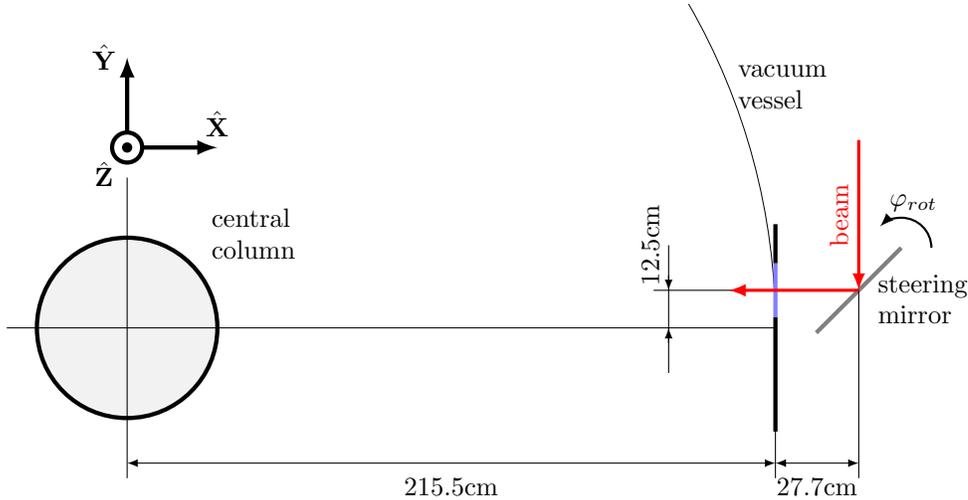
\begin{figure*}
	\centering
	
	\begin{tikzpicture}[>=latex, scale = 0.04]
	
	
	\filldraw[color=black, fill=black!5, ultra thick](0,0) circle (30);
	\node[above right, black, text width=2cm] at (25,20) {central column};
	\draw[black, ultra thin] (215.5,0) arc (0:30:215.5);
	\node[above right, black, text width=2cm] at (200,70) {vacuum vessel};
	\draw[-, ultra thick, blue!50] (215.5, 3.5) -- (215.5, 21.5);
	\draw[-, ultra thick, black] (215.5, 34.5) -- (215.5, 21.5);
	\draw[-, ultra thick, black] (215.5, -34.5) -- (215.5, 3.5);
	\draw[-, ultra thick, black!50] (243.2 - 0.707*20 , 12.5 - 0.707*20) -- (243.2 + 0.707*20, 12.5 + 0.707*20) node[right, black, yshift=-4.5ex, xshift=-3.0ex, text width=2cm] {steering mirror};				
	
	\draw[-, thin, black] (-40, 0) -- (215.5, 0);
	\draw[-, thin, black] (175, 12.5) -- (243.2, 12.5);	
	\draw[-, thin, black] (180, -5) -- (180, 17.5);	
	\draw[->, thin, black] (180, -15) -- (180, 0);	
	\draw[->, thin, black] (180, 27.5) -- (180, 12.5) node[above, yshift=4ex, rotate=90] {$12.5\textrm{cm}$};	
			
	\draw[-, thin, black] (243.2, 12.5) -- (243.2, -50);	
	\draw[-, thin, black] (215.5, 0) -- (215.5, -50);	
	\draw[-, thin, black] (0, 50) -- (0, -50);	
	
	\draw[->, ultra thick, black] (0,60) -- (30,60) node[above, yshift=0ex] {$\hat{\mathbf{X}}$};
	\draw[->, ultra thick, black] (0,60) -- (0,90)  node[left, yshift=0ex] {$\hat{\mathbf{Y}}$};
	\filldraw[color=black, fill=white, ultra thick] (0,60) circle (5);
	\filldraw[color=black, fill=black, ultra thick] (0,60) circle (1) node[below left, xshift=-0.25ex, yshift=-0.25ex] {$\hat{\mathbf{Z}}$};
 				
	\draw[<->, thin, black] (215.5, -45) -- (243.2, -45) node[midway, below, yshift=-0.5ex] { $27.7\textrm{cm}$};	
	\draw[<->, thin, black] (0, -45) -- (215.5, -45) node[midway, below, yshift=-0.5ex] { $215.5\textrm{cm}$};	
					
	\draw[->, black, thick] (243.2 + 0.707*20+10, 12.5 + 0.707*20) arc (0:135:10) node[above right, yshift=0.5ex] {$\varphi_{rot}$};
						
	\draw[->, very thick, red] (243.2, 62.5) -- (243.2, 12.5) node[midway, above,rotate=90] {beam};
	\draw[->, very thick, red] (243.2, 12.5) -- (200, 12.5);

	\end{tikzpicture}

	\caption{Top-down view (toroidal plane) of the previous DBS system on MAST. The steering mirror rotates around two axes, both of which pass through the point where the centre of the beam is incident on its surface. The angle $\varphi_{rot}$ refers to the launch angle in the plane of the diagram, and $\varphi_{tilt}$ adjusts the mirror rotation angle out of plane of the diagram. That is, they are related to the azimuthal and polar angles respectively. A positive $\varphi_{tilt}$ corresponds to mirror's normal pointing towards the sky (and negative corresponds to pointing towards the ground). The sign convention for $\varphi_{rot}$ is given in the diagram. In the diagram, we have $\varphi_{rot} = 0$ and $\varphi_{tilt} = 0$, such that the beam is incident at $45^\circ$, reflected at $45^\circ$, and propagates in the midplane of the tokamak. Note that the probe beam's central ray is offset from intersecting the central column due to the position of the vacuum window. This was taken into account in the geometries used in the calculations.}
	\label{fig:launch_geometry}
	\end{figure*}

	The MAST DBS had two horns, one for the Q-band and the other for the V-band. Both of these horns had their own focusing lenses shown in Figure \ref{fig:horn_geometry}. 
	After a polariser \cite{Hillesheim:DBS_MAST:2015}, the beams were incident, at the same angle, on a steering mirror. This geometry in shown in Figure \ref{fig:horn_geometry}. This mirror was used to control the poloidal and toroidal launch angles of the beam. It is important to realise that the mirror angles, which we call rotation $\varphi_{rot}$ (left and right) and tilt $\varphi_{tilt}$ (up and down), are related to the Scotty (as well as other codes, like Torbeam and Genray) launch angles by a 3D transformation.  A positive $\varphi_{tilt}$ corresponds to the normal of the mirror pointing towards the ceiling (and negative corresponds to pointing towards the ground). The sign convention for $\varphi_{rot}$ is given in Figure \ref{fig:launch_geometry}. When $\varphi_{rot} = 0$ and $\varphi_{tilt} = 0$, the beam is incident on the mirror at $45^\circ$ and reflected at $45^\circ$, subsequently propagating in the midplane of the tokamak. However, $\varphi_{rot} = 0$ does not correspond to the beam being launched in the radial direction (straight toward the central column); in this situation, the beam does not propagate towards the centre of the central column, but is displaced to the side by $12.5$ cm as shown in Figure \ref{fig:launch_geometry}. Those wishing to analyse data from the MAST DBS would do well to take meticulous care to ensure that all the above issues are properly accounted for. In the laboratory Cartesian system ($X$, $Y$, $Z$), the mirror normal is given by
	\begin{equation}
	\fl
	\eqalign{
		& \mathbf{\hat{n}}_{mirror} =
		\left( \begin{array}{ccc}
			\cos \varphi_{rot}' & - \sin \varphi_{rot}' & 0 \\
			\sin \varphi_{rot}' & \cos \varphi_{rot}' & 0 \\
			0		  & 0		 & 1 \\
		\end{array} \right) \\
		& \times \left[
		\left( \begin{array}{ccc}
			1		  & 0		 & 0 \\
			0		  & \cos \varphi_{tilt} & - \sin \varphi_{tilt} \\
			0		  & \sin \varphi_{tilt} & \cos \varphi_{tilt} \\
		\end{array} \right)	
		\left( \begin{array}{c}
			0    \\
			1 \\
			0 \\
		\end{array} \right)	
		\right] , }
	\end{equation}
	where we use the shorthand
	\begin{equation}
		\varphi_{rot}' = \frac{\pi}{4} + \varphi_{rot} - \tan^{-1} {\frac{2155}{125}}.
	\end{equation}
	The arctangent accounts for an angular offset, which is illustrated in Figure \ref{fig:launch_geometry}. With the normal vector of the mirror, the direction of the probe beam as it propagates away from the mirror is calculated, giving the poloidal and toroidal launch angles required as input for Scotty. The initial wavevector in the laboratory frame $(R, \zeta, Z)$ is given by
	\begin{equation} \label{eq:launch_K}
	\eqalign{
		K_{R,ant}     &= - \frac{\Omega}{c}         \cos \varphi_{tor} \cos \varphi_{pol} , \\
		K_{\zeta,ant} &= - \frac{\Omega}{c} R_{ant} \sin \varphi_{tor} \cos \varphi_{pol} , \\
		K_{Z,ant}     &= - \frac{\Omega}{c}       				   \sin \varphi_{pol} ,
	}
	\end{equation}
	where $\varphi_{pol}$ and $\varphi_{tor}$ are the poloidal and toroidal launch angles, respectively. Calculation of the initial beam widths and curvatures for both bands is detailed in \ref{appendix:beam_parameters}.
	
	The hardware was such that the polarisations of the beams from the two horns of the MAST DBS were always perpendicular to each other. These polarisations were optimised such that one horn would launch an O-mode beam, and the other an X-mode beam. Here the O-mode refers to the polarisation being parallel to the magnetic field upon entering the plasma, and the X mode refers to the polarisation being perpendicular to the magnetic field upon entry.
	


	\subsection{Spectral analysis of DBS signal} \label{subsection:DBS_analysis}
	
	We use Welch's method of averaged periodograms \cite{Welch:periodogram:1967} to calculate the spectral density, weighted with Hann windows \cite{Harris:windows:1978} and using a $50\%$ overlap. This method involves dividing a time period into equal segments, applying a Hann window to each of them, individually Fourier transforming them, and then finding the mean spectral density at every given frequency. The standard deviation is then used to estimate the associated standard error of the mean, which we use for error bars. In this work, we average over $101$ segments. Each segment spans $0.1024$ ms, which gives us a total of $5.2$ ms, since any two adjacent segments overlap. We then smooth the spectrum. The smoothing process works as follows. We consider a moving window of 5 points, discard the highest and lowest points, and then take the mean of the remaining 3 points. This is essentially a moving trimmed average.
	
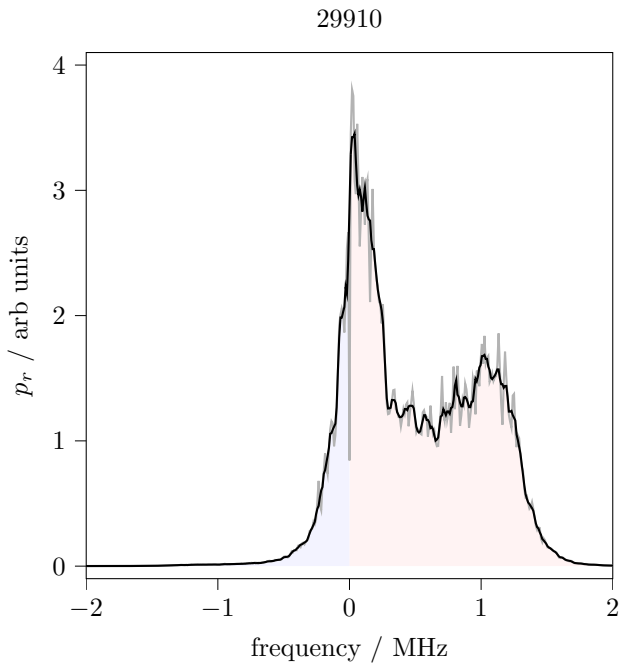
\begin{figure}
	\centering
	\begin{tikzpicture}
	
	\begin{axis}[
	tick align=outside,
	tick pos=left,
	title={29910},
	x grid style={white!69.0196078431373!black},
	xlabel={frequency / MHz},
	xmin=-2, xmax=2,
	xtick style={color=black},
	y grid style={white!69.0196078431373!black},
	ylabel={$p_r$ / arb units},
	ymin=-0.1, ymax=4.1,
	ytick style={color=black},
	width = 7.0cm,
	height=7.0cm,	
	scale only axis
	]
	\addplot [thick, black!30]
	table {%
	-2.001953125 0.000832200050354004
	-1.474609375 0.00287091732025146
	-1.34765625 0.00583386421203613
	-1.337890625 0.00686383247375488
	-1.318359375 0.00744414329528809
	-1.30859375 0.0073704719543457
	-1.279296875 0.010261058807373
	-1.25 0.00748002529144287
	-1.240234375 0.0103358030319214
	-1.23046875 0.00758981704711914
	-1.220703125 0.0101583003997803
	-1.2109375 0.00962615013122559
	-1.201171875 0.0129992961883545
	-1.19140625 0.0108458995819092
	-1.162109375 0.0115587711334229
	-1.15234375 0.0135855674743652
	-1.142578125 0.0107502937316895
	-1.1328125 0.0121220350265503
	-1.123046875 0.0111672878265381
	-1.11328125 0.0125705003738403
	-1.09375 0.0106238126754761
	-1.083984375 0.0110377073287964
	-1.07421875 0.0133210420608521
	-1.064453125 0.0137313604354858
	-1.044921875 0.0114997625350952
	-1.025390625 0.0138025283813477
	-1.015625 0.0136250257492065
	-1.005859375 0.0105854272842407
	-0.986328125 0.0117884874343872
	-0.9765625 0.0156595706939697
	-0.95703125 0.0132983922958374
	-0.947265625 0.018962025642395
	-0.927734375 0.0145069360733032
	-0.91796875 0.0149857997894287
	-0.908203125 0.0138417482376099
	-0.8984375 0.0162564516067505
	-0.87890625 0.0167441368103027
	-0.869140625 0.0168095827102661
	-0.859375 0.0181025266647339
	-0.830078125 0.0179902315139771
	-0.8203125 0.0209383964538574
	-0.810546875 0.0172404050827026
	-0.80078125 0.0176379680633545
	-0.791015625 0.0210143327713013
	-0.78125 0.0231733322143555
	-0.771484375 0.0201950073242188
	-0.76171875 0.0211436748504639
	-0.751953125 0.0250803232192993
	-0.7421875 0.0267869234085083
	-0.732421875 0.0208132266998291
	-0.712890625 0.0282125473022461
	-0.703125 0.0260722637176514
	-0.693359375 0.0197025537490845
	-0.68359375 0.0237395763397217
	-0.6640625 0.0288972854614258
	-0.654296875 0.0261772871017456
	-0.625 0.0365915298461914
	-0.615234375 0.0412881374359131
	-0.60546875 0.0431718826293945
	-0.595703125 0.0360218286514282
	-0.5859375 0.0381249189376831
	-0.576171875 0.0418211221694946
	-0.56640625 0.0498420000076294
	-0.556640625 0.0465039014816284
	-0.537109375 0.0514105558395386
	-0.52734375 0.0499012470245361
	-0.517578125 0.0532034635543823
	-0.498046875 0.0755552053451538
	-0.48828125 0.0787922143936157
	-0.478515625 0.0720715522766113
	-0.46875 0.0697782039642334
	-0.458984375 0.0761680603027344
	-0.439453125 0.11832332611084
	-0.4296875 0.110112428665161
	-0.419921875 0.128878593444824
	-0.400390625 0.127054691314697
	-0.390625 0.163323760032654
	-0.380859375 0.1677086353302
	-0.37109375 0.153152942657471
	-0.361328125 0.182012796401978
	-0.3515625 0.187492847442627
	-0.341796875 0.173755407333374
	-0.33203125 0.19382905960083
	-0.322265625 0.204184651374817
	-0.3125 0.218661308288574
	-0.302734375 0.279986381530762
	-0.29296875 0.286358714103699
	-0.283203125 0.297936201095581
	-0.2734375 0.355193018913269
	-0.263671875 0.334655165672302
	-0.25390625 0.371595025062561
	-0.244140625 0.515749216079712
	-0.234375 0.680674076080322
	-0.224609375 0.4841228723526
	-0.21484375 0.446923136711121
	-0.205078125 0.650559067726135
	-0.1953125 0.760677099227905
	-0.185546875 0.900826454162598
	-0.17578125 0.797999382019043
	-0.166015625 0.737415194511414
	-0.15625 0.940355658531189
	-0.146484375 1.09164273738861
	-0.13671875 1.15707683563232
	-0.126953125 1.13805890083313
	-0.1171875 0.956826210021973
	-0.107421875 1.04161703586578
	-0.09765625 1.22939240932465
	-0.087890625 1.53628659248352
	-0.078125 1.92633509635925
	-0.068359375 2.00728464126587
	-0.05859375 2.05863308906555
	-0.048828125 2.01968932151794
	-0.0390625 1.86541509628296
	-0.029296875 2.12798476219177
	-0.01953125 2.53590178489685
	-0.009765625 2.66654562950134
	0 0.843818664550781
	0.009765625 3.4470853805542
	0.01953125 3.8039391040802
	0.029296875 3.75347256660461
	0.0390625 3.07163310050964
	0.048828125 2.96974945068359
	0.05859375 3.53078651428223
	0.068359375 3.07878017425537
	0.078125 2.55286955833435
	0.09765625 3.10688233375549
	0.107421875 2.93168473243713
	0.1171875 2.72684264183044
	0.126953125 3.0412003993988
	0.13671875 3.08930611610413
	0.146484375 2.62601041793823
	0.15625 2.10804510116577
	0.166015625 2.63904476165771
	0.17578125 3.01116108894348
	0.185546875 2.53986024856567
	0.205078125 2.28227686882019
	0.21484375 2.23661971092224
	0.224609375 2.10708284378052
	0.234375 2.09512758255005
	0.244140625 1.93776977062225
	0.25390625 2.08932185173035
	0.263671875 1.81647646427155
	0.2734375 1.50772976875305
	0.283203125 1.29739475250244
	0.29296875 1.21322786808014
	0.302734375 1.21781373023987
	0.3125 1.26055002212524
	0.322265625 1.31458389759064
	0.33203125 1.40409362316132
	0.341796875 1.40790104866028
	0.3515625 1.25474321842194
	0.361328125 1.2430956363678
	0.37109375 1.19416439533234
	0.380859375 1.19540429115295
	0.390625 1.11559700965881
	0.400390625 1.17911064624786
	0.41015625 1.25562739372253
	0.419921875 1.2168151140213
	0.4296875 1.28979504108429
	0.439453125 1.30667853355408
	0.44921875 1.11679542064667
	0.458984375 1.22926712036133
	0.46875 1.30646824836731
	0.478515625 1.40050089359283
	0.48828125 1.30355930328369
	0.498046875 1.16827785968781
	0.5078125 1.06789565086365
	0.517578125 1.09454047679901
	0.52734375 1.07415747642517
	0.537109375 1.05877590179443
	0.546875 1.05155873298645
	0.556640625 1.22743225097656
	0.56640625 1.24364638328552
	0.576171875 1.19539046287537
	0.5859375 1.19169783592224
	0.595703125 1.03428757190704
	0.60546875 1.11419153213501
	0.615234375 1.28115522861481
	0.625 1.14007174968719
	0.634765625 1.05445551872253
	0.64453125 1.08207190036774
	0.654296875 0.990570545196533
	0.6640625 0.957191586494446
	0.673828125 0.95947527885437
	0.68359375 1.1317800283432
	0.693359375 1.23365569114685
	0.703125 1.51816880702972
	0.712890625 1.21635258197784
	0.72265625 1.16079914569855
	0.732421875 1.30900394916534
	0.7421875 1.16889882087708
	0.751953125 1.23234415054321
	0.76171875 1.3036208152771
	0.771484375 1.06410646438599
	0.78125 1.25102305412292
	0.791015625 1.5926787853241
	0.80078125 1.25320518016815
	0.810546875 1.47388362884521
	0.8203125 1.5985940694809
	0.830078125 1.37912273406982
	0.83984375 1.11923611164093
	0.849609375 1.20242834091187
	0.859375 1.33298444747925
	0.869140625 1.29150307178497
	0.87890625 1.45666587352753
	0.888671875 1.41102874279022
	0.8984375 1.30653119087219
	0.908203125 1.10534071922302
	0.927734375 1.29959094524384
	0.9375 1.49833452701569
	0.947265625 1.43603146076202
	0.95703125 1.21214115619659
	0.966796875 1.55550229549408
	0.9765625 1.5601714849472
	0.986328125 1.43627905845642
	0.99609375 1.63936054706573
	1.005859375 1.77388429641724
	1.015625 1.61564993858337
	1.025390625 1.83817028999329
	1.03515625 1.6205803155899
	1.044921875 1.66329419612885
	1.0546875 1.67844724655151
	1.064453125 1.4693888425827
	1.07421875 1.49575805664062
	1.083984375 1.53824388980865
	1.09375 1.47823143005371
	1.103515625 1.50620937347412
	1.11328125 1.51675844192505
	1.123046875 1.60242569446564
	1.1328125 1.85795855522156
	1.15234375 1.30082678794861
	1.162109375 1.1291116476059
	1.171875 1.46469712257385
	1.181640625 1.71181166172028
	1.19140625 1.50635600090027
	1.201171875 1.34387636184692
	1.2109375 1.2063854932785
	1.220703125 1.15084517002106
	1.23046875 1.10639357566833
	1.240234375 1.36100733280182
	1.25 1.374387383461
	1.259765625 1.08154463768005
	1.26953125 0.949637889862061
	1.279296875 1.04493927955627
	1.2890625 1.0221426486969
	1.298828125 0.875776767730713
	1.30859375 0.792294263839722
	1.318359375 0.762043237686157
	1.328125 0.614120244979858
	1.337890625 0.549619436264038
	1.34765625 0.559513807296753
	1.357421875 0.503192901611328
	1.3671875 0.458609104156494
	1.376953125 0.472367525100708
	1.38671875 0.510427832603455
	1.396484375 0.457129716873169
	1.40625 0.393067598342896
	1.416015625 0.320780515670776
	1.42578125 0.288308262825012
	1.435546875 0.308014988899231
	1.4453125 0.304998278617859
	1.455078125 0.218906044960022
	1.46484375 0.190608859062195
	1.474609375 0.222725629806519
	1.484375 0.208390474319458
	1.494140625 0.177321195602417
	1.50390625 0.150384426116943
	1.513671875 0.153019189834595
	1.5234375 0.142534613609314
	1.533203125 0.139562368392944
	1.54296875 0.125308990478516
	1.572265625 0.109115719795227
	1.591796875 0.068820595741272
	1.6015625 0.069642186164856
	1.611328125 0.0717484951019287
	1.62109375 0.0690343379974365
	1.630859375 0.0571552515029907
	1.640625 0.0482747554779053
	1.650390625 0.0490198135375977
	1.669921875 0.0379874706268311
	1.6796875 0.0424973964691162
	1.689453125 0.04114830493927
	1.69921875 0.0290741920471191
	1.708984375 0.023242712020874
	1.71875 0.0231328010559082
	1.728515625 0.0255593061447144
	1.73828125 0.0208238363265991
	1.7578125 0.0216187238693237
	1.767578125 0.0201873779296875
	1.77734375 0.0156726837158203
	1.787109375 0.0155305862426758
	1.796875 0.0177066326141357
	1.806640625 0.0152835845947266
	1.81640625 0.0113935470581055
	1.8359375 0.0117590427398682
	1.85546875 0.0120465755462646
	1.865234375 0.0105172395706177
	1.875 0.0108288526535034
	1.923828125 0.00556027889251709
	1.93359375 0.00627171993255615
	1.982421875 0.00391745567321777
	2.001953125 0.00467145442962646
	};
	\addplot [thick, black, name path = negsignal]
	table {%
	-2.001953125 0.000832080841064453
	-1.66015625 0.00104200839996338
	-1.455078125 0.00307011604309082
	-1.30859375 0.00778567790985107
	-1.279296875 0.00901806354522705
	-1.23046875 0.009124755859375
	-1.2109375 0.0102100372314453
	-1.171875 0.0115929841995239
	-1.064453125 0.0123195648193359
	-1.03515625 0.0129059553146362
	-0.986328125 0.012480616569519
	-0.966796875 0.0142999887466431
	-0.95703125 0.0154992341995239
	-0.9375 0.0154629945755005
	-0.888671875 0.0163565874099731
	-0.810546875 0.0188555717468262
	-0.80078125 0.0198636054992676
	-0.791015625 0.0196157693862915
	-0.76171875 0.0231324434280396
	-0.751953125 0.0223457813262939
	-0.72265625 0.0256377458572388
	-0.712890625 0.0236465930938721
	-0.693359375 0.0254055261611938
	-0.673828125 0.0254405736923218
	-0.634765625 0.0334382057189941
	-0.625 0.0373163223266602
	-0.60546875 0.0386682748794556
	-0.5859375 0.0410392284393311
	-0.576171875 0.0421500205993652
	-0.556640625 0.0483458042144775
	-0.52734375 0.0515050888061523
	-0.517578125 0.0562450885772705
	-0.5078125 0.0642932653427124
	-0.498046875 0.0705826282501221
	-0.46875 0.0756772756576538
	-0.458984375 0.0815554857254028
	-0.4296875 0.118735551834106
	-0.419921875 0.124382972717285
	-0.41015625 0.127901315689087
	-0.400390625 0.139991044998169
	-0.390625 0.148082494735718
	-0.380859375 0.161395072937012
	-0.37109375 0.171015024185181
	-0.361328125 0.174492239952087
	-0.33203125 0.195168852806091
	-0.322265625 0.205558300018311
	-0.302734375 0.261668801307678
	-0.29296875 0.28809380531311
	-0.283203125 0.306316733360291
	-0.263671875 0.353814363479614
	-0.25390625 0.414179086685181
	-0.244140625 0.457155704498291
	-0.234375 0.482265114784241
	-0.224609375 0.550143718719482
	-0.21484375 0.605118751525879
	-0.205078125 0.631786346435547
	-0.1953125 0.736411809921265
	-0.185546875 0.765363931655884
	-0.166015625 0.879727125167847
	-0.15625 0.943332672119141
	-0.146484375 1.05668580532074
	-0.13671875 1.06217586994171
	-0.126953125 1.09043955802917
	-0.107421875 1.13635611534119
	-0.09765625 1.26909863948822
	-0.087890625 1.56400465965271
	-0.078125 1.82330214977264
	-0.068359375 1.98443639278412
	-0.05859375 1.98443639278412
	-0.0390625 2.0687689781189
	-0.029296875 2.227858543396
	-0.01953125 2.176433801651
	-0.009765625 2.44347739219666
	0.009765625 3.28903460502625
	};
	\addplot [thick, black, name path = possignal]
	table {%
	0.009765625 3.28903460502625
	0.01953125 3.42406368255615
	0.029296875 3.42406368255615
	0.0390625 3.45196413993835
	0.048828125 3.2270667552948
	0.05859375 3.04005432128906
	0.068359375 2.96067881584167
	0.078125 3.00638961791992
	0.087890625 2.94799041748047
	0.09765625 2.83067798614502
	0.107421875 2.93546390533447
	0.1171875 3.02073049545288
	0.126953125 2.89990925788879
	0.13671875 2.79801774024963
	0.146484375 2.76875185966492
	0.15625 2.75873875617981
	0.166015625 2.60163855552673
	0.17578125 2.53088712692261
	0.185546875 2.53088712692261
	0.1953125 2.41196465492249
	0.21484375 2.20865988731384
	0.224609375 2.1462767124176
	0.234375 2.09717750549316
	0.244140625 2.04073977470398
	0.25390625 1.94785606861115
	0.2734375 1.54053366184235
	0.283203125 1.34097945690155
	0.29296875 1.25858616828918
	0.302734375 1.25858616828918
	0.3125 1.26431584358215
	0.322265625 1.3264092206955
	0.33203125 1.3264092206955
	0.341796875 1.32447361946106
	0.3515625 1.30064415931702
	0.361328125 1.23108100891113
	0.380859375 1.18955969810486
	0.390625 1.18955969810486
	0.400390625 1.19711005687714
	0.41015625 1.21718430519104
	0.419921875 1.25407910346985
	0.4296875 1.25407910346985
	0.439453125 1.24529242515564
	0.44921875 1.27517676353455
	0.458984375 1.28080463409424
	0.478515625 1.27976489067078
	0.48828125 1.25943505764008
	0.5078125 1.11232531070709
	0.517578125 1.07886457443237
	0.52734375 1.06694304943085
	0.537109375 1.07582461833954
	0.56640625 1.20484018325806
	0.576171875 1.20484018325806
	0.5859375 1.16709327697754
	0.595703125 1.16709327697754
	0.60546875 1.14865374565125
	0.615234375 1.10290622711182
	0.625 1.11211168766022
	0.634765625 1.09219980239868
	0.64453125 1.04236602783203
	0.654296875 1.00150048732758
	0.6640625 1.01070594787598
	0.673828125 1.02727520465851
	0.693359375 1.19392943382263
	0.703125 1.20360243320465
	0.712890625 1.25300407409668
	0.72265625 1.23141849040985
	0.732421875 1.20586514472961
	0.7421875 1.2349545955658
	0.751953125 1.2349545955658
	0.76171875 1.2174220085144
	0.771484375 1.26232933998108
	0.78125 1.26928305625916
	0.791015625 1.3260372877121
	0.80078125 1.43992257118225
	0.810546875 1.48189508914948
	0.8203125 1.36873722076416
	0.830078125 1.35181152820587
	0.83984375 1.30484521389008
	0.849609375 1.27563858032227
	0.859375 1.27563858032227
	0.869140625 1.34517204761505
	0.87890625 1.35018146038055
	0.888671875 1.33635437488556
	0.8984375 1.3062344789505
	0.908203125 1.26908850669861
	0.91796875 1.26908850669861
	0.927734375 1.31225526332855
	0.9375 1.31592118740082
	0.95703125 1.49662280082703
	0.966796875 1.47593760490417
	0.9765625 1.51731765270233
	0.986328125 1.58501148223877
	0.99609375 1.60506069660187
	1.005859375 1.67629826068878
	1.015625 1.67794167995453
	1.025390625 1.68591952323914
	1.03515625 1.65410721302032
	1.044921875 1.65410721302032
	1.0546875 1.59321081638336
	1.064453125 1.56576538085938
	1.07421875 1.50407779216766
	1.083984375 1.49339962005615
	1.103515625 1.52040386199951
	1.11328125 1.54179787635803
	1.123046875 1.57047998905182
	1.1328125 1.57047998905182
	1.142578125 1.49850273132324
	1.15234375 1.4525933265686
	1.162109375 1.4525933265686
	1.171875 1.42395997047424
	1.181640625 1.43830978870392
	1.19140625 1.43830978870392
	1.201171875 1.352205991745
	1.2109375 1.23370230197906
	1.220703125 1.23370230197906
	1.23046875 1.23941266536713
	1.240234375 1.2060821056366
	1.259765625 1.16249704360962
	1.26953125 1.04954218864441
	1.279296875 1.00557327270508
	1.298828125 0.896737933158875
	1.328125 0.645225763320923
	1.337890625 0.57441782951355
	1.34765625 0.537441968917847
	1.357421875 0.508393287658691
	1.3671875 0.4953293800354
	1.38671875 0.46270215511322
	1.396484375 0.440855026245117
	1.416015625 0.340620994567871
	1.42578125 0.311264634132385
	1.435546875 0.300440549850464
	1.4453125 0.270737528800964
	1.455078125 0.248876690864563
	1.46484375 0.216674089431763
	1.474609375 0.205968499183655
	1.494140625 0.179576873779297
	1.50390625 0.160241603851318
	1.513671875 0.148646116256714
	1.5234375 0.144160509109497
	1.552734375 0.119226455688477
	1.5625 0.113828659057617
	1.572265625 0.103655457496643
	1.591796875 0.0768395662307739
	1.6015625 0.0701416730880737
	1.611328125 0.0691657066345215
	1.62109375 0.0652772188186646
	1.640625 0.0514832735061646
	1.650390625 0.0468541383743286
	1.6796875 0.0405443906784058
	1.708984375 0.025958776473999
	1.728515625 0.0226644277572632
	1.7578125 0.0208762884140015
	1.787109375 0.0163033008575439
	1.81640625 0.0128120183944702
	1.8359375 0.0117330551147461
	1.865234375 0.0111309289932251
	1.93359375 0.0060117244720459
	2.001953125 0.0042414665222168
	};
    \path[name path=axis] (axis cs:-2,0) -- (axis cs:2,0);
	\addplot [
		thick,
		color=blue,
		fill=blue, 
		fill opacity=0.05
	]
		fill between[
		of=
		negsignal and axis,
		soft clip={domain=-2:0},
	];		
	\addplot [
		thick,
		color=red,
		fill=red, 
		fill opacity=0.05
	]
		fill between[
		of=
		possignal and axis,
		soft clip={domain=0:2},
	];		
	\end{axis}
	\end{tikzpicture}
	\caption{Welch periodogram (grey line) of the spectral density and the smoothed spectrum (black line). We calculate the backscattered power by subtracting the blue area from the red area, which is a good approximation as the zero-frequency peaks in this study were sufficiently symmetric about $f=0$.}
	\label{fig:DBS_data_analysis}
	\end{figure}

	DBS spectra are often formed of two peaks: one centred at zero frequency, with the other at some Doppler-shifted frequency, as shown in Figure \ref{fig:DBS_data_analysis}. The latter is believed to contain the useful backscattered signal, while the former is thought to be a spurious contribution \cite{Happel:DBS:2010}. Hence, one tries to remove the zero-frequency peak, and analyse the Doppler-shifted peak. This is typically done by fitting peaks to the spectrum, enabling one to find the Doppler shift and backscattered power of the signal. The Doppler shift is generally of interest as it sheds light on the velocity profile of the plasma. However, in this study we are not interested in the frequency of this shift, and fitting the data is difficult because the MAST DBS Doppler-shifted peak is often asymmetric and broad; instead, we focus on the backscattered power. Consequently, the analysis procedure may be significantly simplified compared to that of previous work \cite{Hillesheim:DBS_MAST:2015}. We find the total backscattered power at both positive and negative frequencies and subtract the latter from the former, getting what we call the antisymmetric backscattered power, see Figure \ref{fig:DBS_data_analysis}. Using the antisymmetric backscattered power for DBS analysis is a technique also used by other groups \cite{Wen:DBS:2019}. This approach relies on the zero-frequency peak being symmetric about zero frequency. Different techniques for finding the backscattered power were also explored; they turned out to be as robust for the purposes of this work. Hence, we chose the simplest method, the one described above. As for the MAST-U DBS, we fitted the Doppler shifted peak with a Gaussian and calculated the power accordingly.
		
	\section{Mismatch attenuation} \label{section:mismatch}
	Having discussed our model, shots studied, experimental setup, and data analysis approach in the previous section, we now explain how they come together to quantitatively elucidate the effect of mismatch on DBS signals. We first describe how we use the mismatch factor, $F_m$, and the other instrumentation weighting functions, $F_r$, $F_b$, and $F_p$, to evaluate the variation of backscattered power with toroidal launch angle, subsection \ref{subsection:weighting}. We then present comparison with O-mode MAST V-band data in subsection \ref{subsection:data_V} and with X-mode MAST Q-band data in subsection \ref{subsection:data_Q}.

	\subsection{Evaluating contributions to the backscattered signal} \label{subsection:weighting}
	We now describe how we bring together the beam model's insight on instrumentation weighting and experiments to calculate the effect of mismatch attenuation. 
	
\begin{figure}
		\centering	
		\includegraphics[width=0.7\columnwidth]{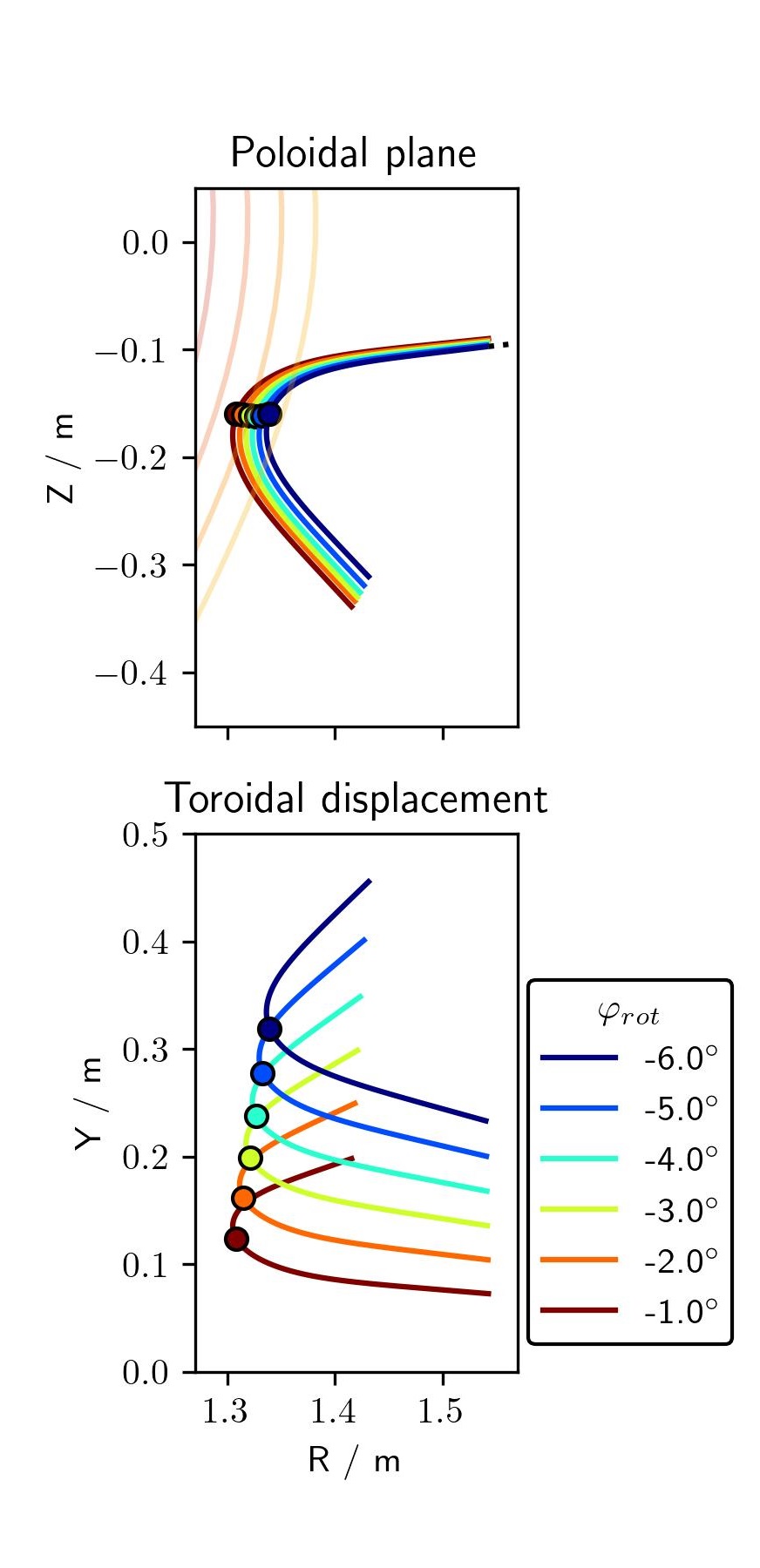}
		\caption{Trajectories of the 55 GHz MAST DBS probe beam at 240 ms. These beams are launched with different toroidal angles, corresponding to the angles in shot group 1, and reach the cut-off at similar flux surfaces. For each ray, the circle marks the cut-off location, defined to be the point where the wavenumber $K$ is minimised.}
		\label{fig:all_angles}
		\end{figure}

\begin{figure}
		\centering
		\includegraphics[width=0.99\columnwidth]{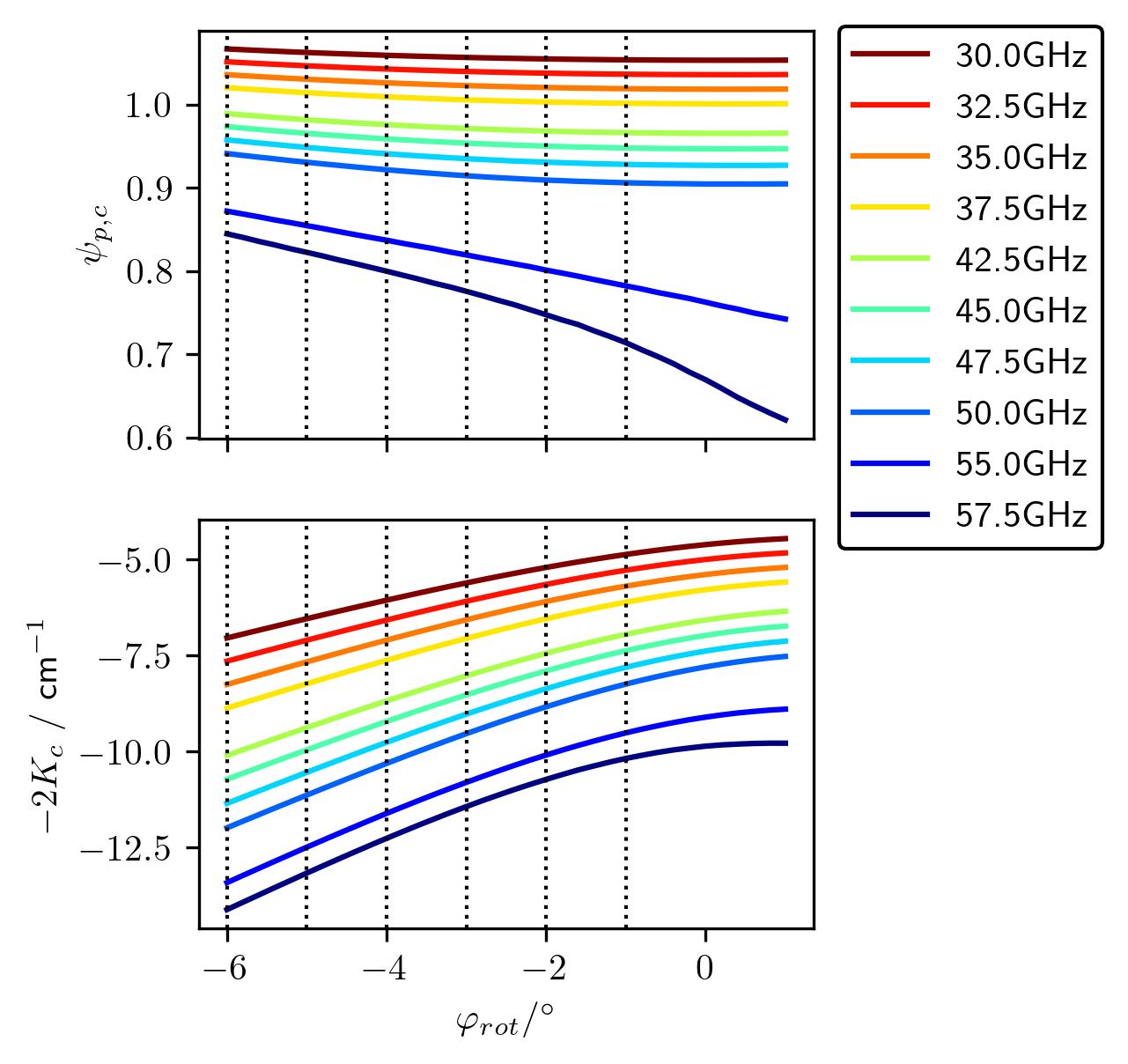}
		\caption{The normalised flux surface label at cut-off, $\varphi_{p,c}$, (top) and backscattered wavenumber (bottom) at the cut-off for different frequencies and launch angles, for MAST shot group 1 at 240 ms. The 55 GHz line summarises Figure \ref{fig:all_angles}. The experimental mirror rotation angles are given by the vertical dotted lines; they run from $-6^{\circ}$ to $-1^{\circ}$, as shown in Table \ref{table:shotgroup}. Only channels with frequencies below the maximum cut-off frequencies, see Figure \ref{fig:equilibria}, are plotted.} 
		\label{fig:cutoff_polflux}
	\end{figure}

	
	Experimental investigation of the mismatch attenuation factor at the cut-off, $F_{m,c}$, is usually \cite{Hillesheim:DBS_MAST:2015, Hall-Chen:beam_model_DBS:2022, Hall-Chen:mismatch:2022, Damba:mismatch:2021, Damba:mismatch:2022} done as follows. The poloidal launch angle is fixed and the toroidal launch angle varied with each repeated shot, Figure \ref{fig:all_angles}. We aim for significant variation in only the mismatch angle. As such, the difference in backscattered powers for any given channel can then typically be attributed entirely to the mismatch attenuation.
	Quite unfortunately, for shot group 1, we find that there is indeed a significant variation in the cut-off location and the corresponding wavenumber at that location as the toroidal launch angle is swept, Figure \ref{fig:cutoff_polflux}. This is rather unlike the toroidal scan on DIII-D \cite{Damba:mismatch:2022, Hall-Chen:mismatch:2022}, where such a variation was minimal.
	Taking the 57.5 GHz beam as an example, this variation corresponds to a change of the cut-off location from a normalised poloidal flux of 0.84 to 0.71 and thus a beam wavenumber from $10 \ \textrm{cm}^{-1}$ to $14 \ \textrm{cm}^{-1}$; a significant variation. Drawing insight by analogy with high-k scattering \cite{Slusher:scattering:1980}, the scattered signal should decrease with increasing wavenumber as the scattering efficiency is lower. As a result of the Bragg condition and the power-law spectrum of turbulence, we would also expected the backscattered signal to decrease with increasing wavenumber. Moreover, the statistical properties of turbulence might change with location; for example, the exponent in the power law could be different. To avoid these challenges in future experiments, instead of effectively fixing the poloidal angle, one could carefully vary the poloidal angle during the toroidal sweep to counter this change, such that the wavenumber at cut-off and cut-off location stay similar. This would require another set of repeated shots with meticulous planning to optimise the DBS setup.

\begin{figure*}
	\centering	
	\includegraphics[width=0.75\textwidth]{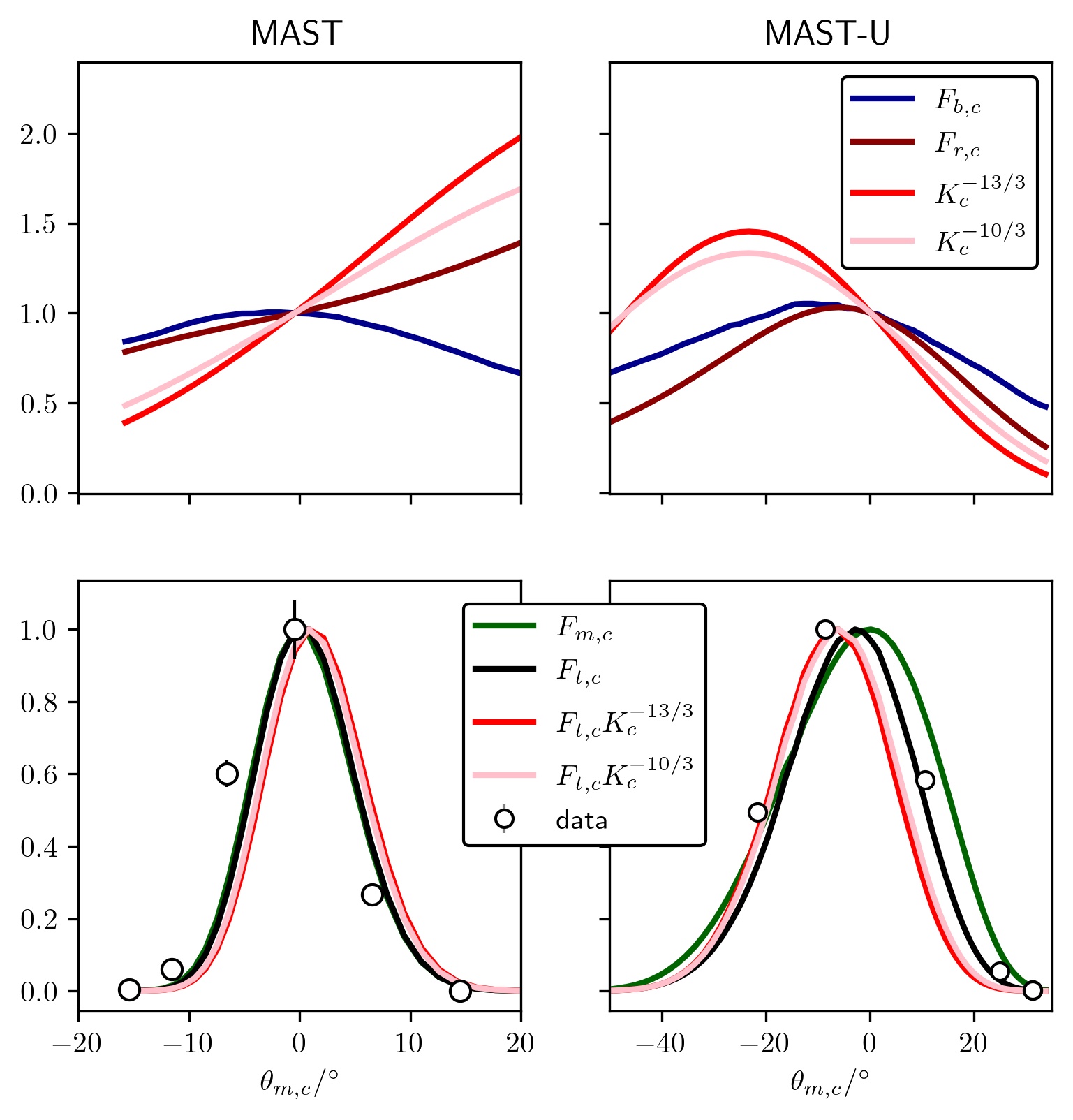}
	\caption{Instrumentation weights at the cut-off as a function of mirror rotation angle (MAST) and toroidal launch angle (MAST-U). We show results from the 57.5 GHz channel of the MAST DBS in O-mode (shot group 1 at 240 ms) and the 32.5 GHz channel of the MAST-U DBS in X-mode. Here the subscript $_c$ indicates evaluation of the quantity at the nominal cut-off location, that is, the point along the central ray where the beam's wavenumber is minimised. The polarisation piece $F_{p,c}$ is included in the calculation of $F_{t,c}$, but it is not plotted separately as it does not vary significantly. To account for the power-law turbulence spectrum, we use the pieces $K_c^{-10/3}$ and $K_c^{-13/3}$. The various pieces are normalised to their respective values at zero mismatch. We see that the change in backscattered power is mostly due to the mismatch attenuation; the other pieces may change significantly (top row), but ultimately have only a small effect on both the width and peak of the predicted toroidal variation in backscattered signal (bottom row). We plot these widths, which we call mismatch tolerance, in Figure \ref{fig:MAST_Vband}, showing that the other instrumentation weights and turbulence spectra affected the decay half width by under a degree. This effect is so small that the different lines overlap significantly in the bottom two figures.
	}
	\label{fig:localisation_torscan}
	\end{figure*}

	In this work, we adopt an alternative approach. We use our model \cite{Hall-Chen:beam_model_DBS:2022} to quantitatively calculate the effect of different wavenumbers and turbulence spectra on the backscattered signal. In our model, we do not explicitly have a scattering efficiency piece. The magnitude of the group velocity in the ray piece, equation (\ref{eq:ray_factor}), is proportional to wavenumber in slab geometry \cite{RuizRuiz:slab:2024}. However, the physical intuition of this piece is not that of scattering efficiency, but rather, the amplitude of the DBS probe beam's electric field. This ray piece arises from the $g^{- 1 / 2}$ in the amplitude of the probe beam's electric field, equation (\ref{eq:beam_field_final}). To study how the turbulence spectrum affects the backscattered power, we take $\tilde{C} \propto k_\perp^{-\alpha} \propto K_c^{-\alpha}$ use the exponents $\alpha = 13/3$ and $\alpha = 10/3$.	We find that both instrumentation effects and amplitude of measured turbulence due to the power-law spectrum do indeed vary significantly as the toroidal angle is changed. In particular, $K_c^{-13/3}$ varies by order unity. Rather surprisingly, the beam model found that mismatch piece remained the dominant contribution to the variation in backscattered signal with toroidal angle. While the other instrumentation weights and turbulence spectra did affect both the width and the peak backscattered power as a function of toroidal launch angle, the overall effect was small and, given the uncertainty in our experimental data, not conclusively distinguishable from that by mismatch attenuation alone, Figure \ref{fig:localisation_torscan}. Interestingly, while the turbulence had a noticeable effect, the choice of exponent did not make a significant difference. These results are the same for both the MAST and MAST-U shot studied.
	
	In the next subection, we extend our analysis by studying all channels of the MAST Q and V bands at various times.

	\subsection{Validation: V-band DBS on MAST, O-mode} \label{subsection:data_V}
	This is the first systematic study of quantitatively predicting the mismatch attenuation of O-mode DBS. Previous work only used a single time and a single frequency with a finite toroidal offset \cite{Hall-Chen:beam_model_DBS:2022} or focused on X-mode \cite{Hall-Chen:mismatch:2022, Damba:mismatch:2022}. We limit our analysis to cases where most of the backscattered signal comes from the region around the cut-off. This corresponds to the lowest two frequency channels at later times in the shots, when the electron densities were higher.
	
\begin{figure*}
		\centering
		\begin{subfigure}{0.4\textwidth}		
			\includegraphics[width=\textwidth]{{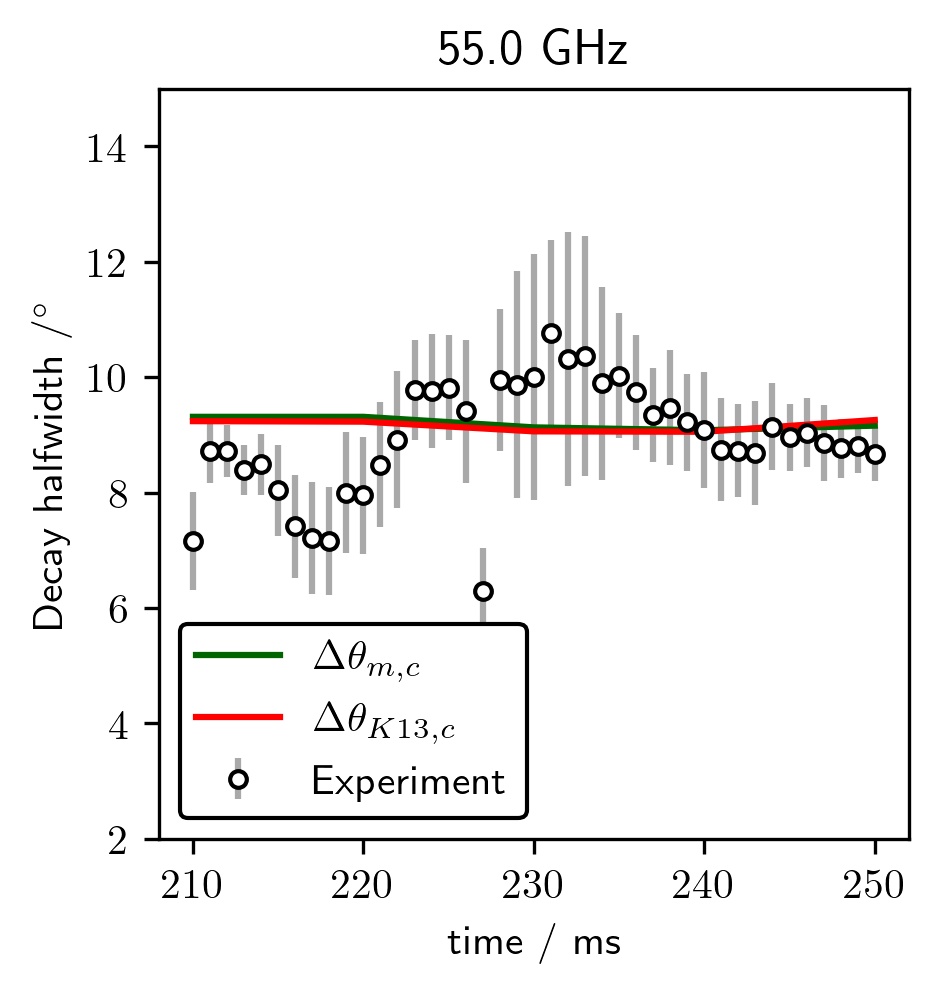}}
		\end{subfigure}	
		\begin{subfigure}{0.4\textwidth}		
			\includegraphics[width=\textwidth]{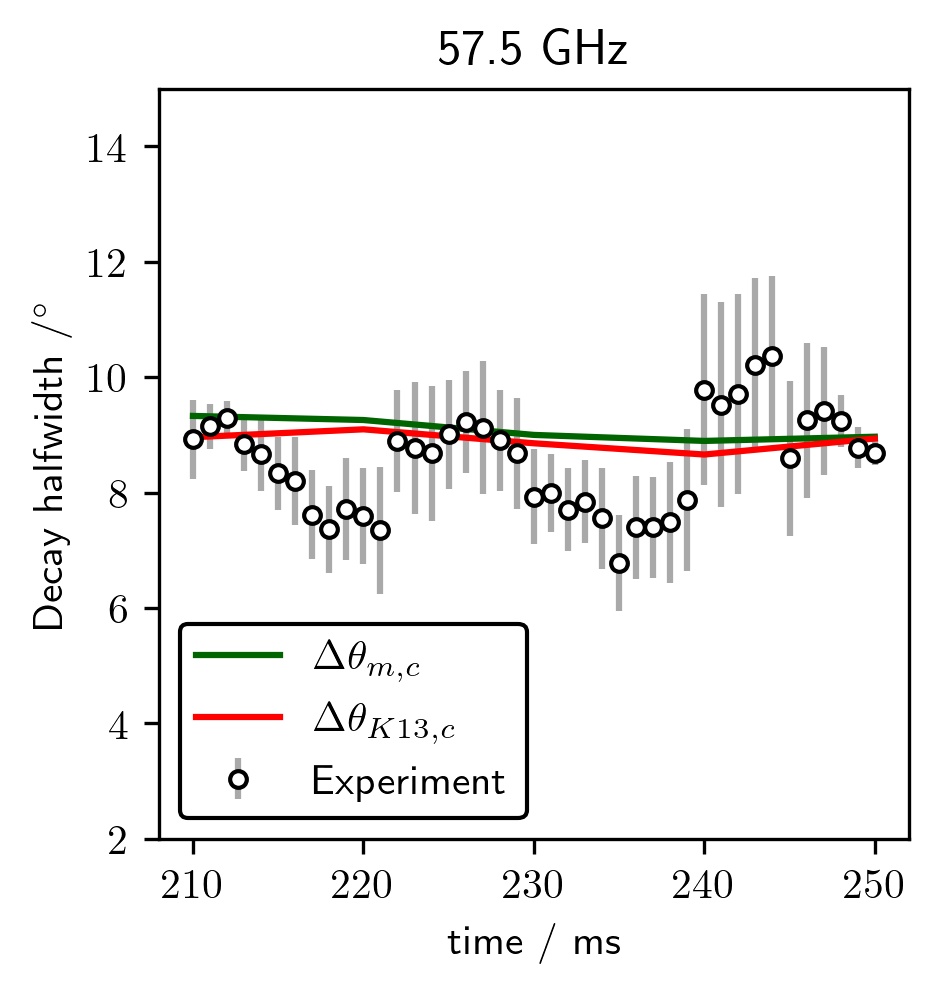}
		\end{subfigure}	
		\caption{The decay halfwidths of backscattered power as a function of time, for shot group 1: Gaussian fits of experimental data (points), the mismatch tolerance factor (green line), and the total instrumentation weight (black line). These were O-mode channels. Calculation of the error bars is described in Section \ref{subsection:DBS_analysis}. As the data are from repeated shots and not simulatenous measurements of the same shot, the error bars and the change in backscattered power should be largely attributed to shot-to-shot variation in plasma properties. Also note that the plasma density is increasing during this time interval, Figure \ref{fig:equilibria}. We see that the other instrumentation weights do not have a significant effect on the apparent mismatch tolerance (red line), despite the variation in cut-off wavenumber and cut-off location, as shown in Figure \ref{fig:cutoff_polflux}.}
		\label{fig:MAST_Vband}
	\end{figure*}

	We find that the mismatch attenuation and total instrumentation weighting functions, both evaluated at the cut-off, predict similar variations of backscattered power as a function of $\varphi_{rot}$. To summarise large amounts of data, we compare the $1/e^2$ halfwidths of the mismatch tolerance. These results are shown in Figure \ref{fig:MAST_Vband}. The beam model's predictions largely agree with experiments; the experimental fit's error bars are large because the toroidal sweep was coarse, in only three of the six repeated shots do the DBS diagnostic measure significant backscattered power. Moreover, there is some small variation in the plasma equilibrium from shot to shot. We consistently find that, despite apparently non-negligible variation in wavenumber at cut-off and cut-off location, its effect on the mismatch tolerance is small.
	Interestingly, we also see that the mismatch tolerance at nominal cut-off, $\Delta \theta_{m,c}$, does not change significantly with time. One can obtain some intuition of this phenomena from considering DBS in slab geometry. In this case, mismatch tolerance depends strongly on the $\Psi_{xx}$ component of the beam \cite{Hall-Chen:beam_model_DBS:2022}; while the beam focuses strongly due to the plasma, this focusing is in the $\Psi_{yy}$ direction \cite{RuizRuiz:slab:2024} instead.
	

	\subsection{Validation: Q-band DBS on MAST, X-mode} \label{subsection:data_Q}
	The beam model's predictions of mismatch attenuation was previously validated on DIII-D plasmas for DBS in X-mode at a single time \cite{Hall-Chen:mismatch:2022, Damba:mismatch:2022}. We now extend this study to spherical tokamak plasmas at different times. 
	
\begin{figure*}
		\centering
		\begin{subfigure}{0.4\textwidth}		
			\includegraphics[width=\textwidth]{{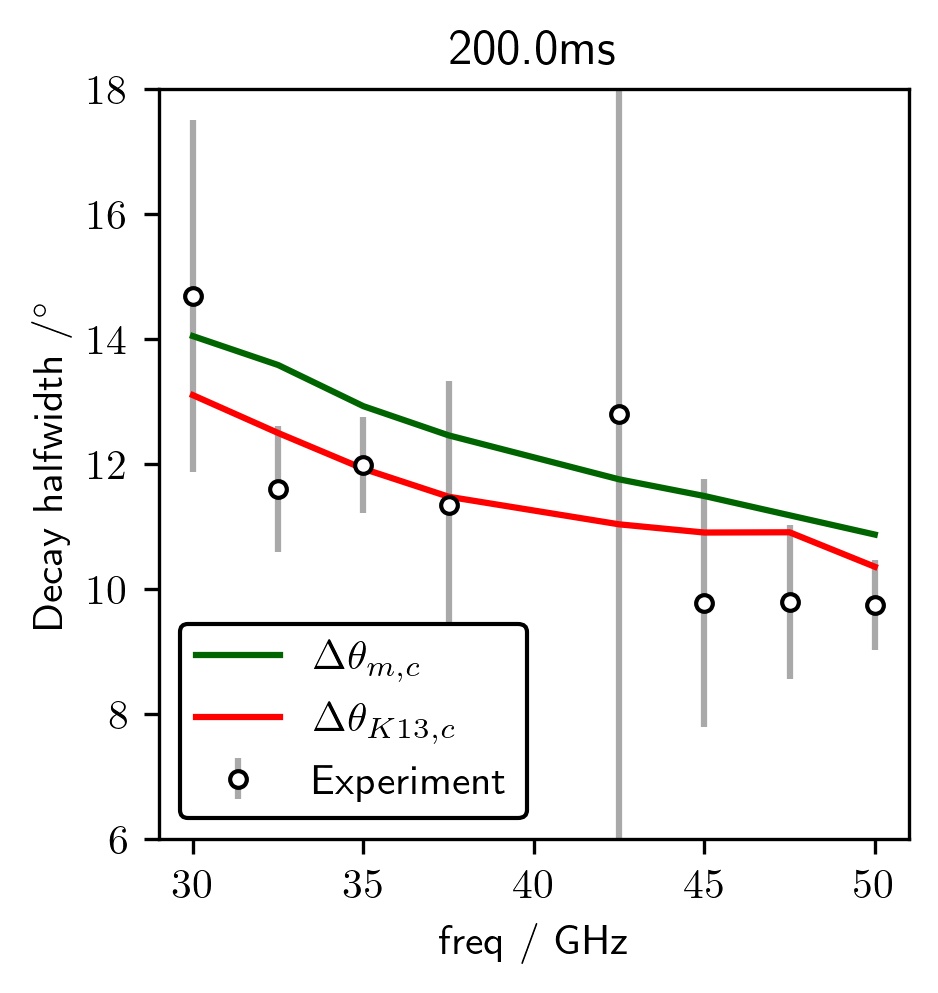}}
		\end{subfigure}	
		\begin{subfigure}{0.4\textwidth}		
			\includegraphics[width=\textwidth]{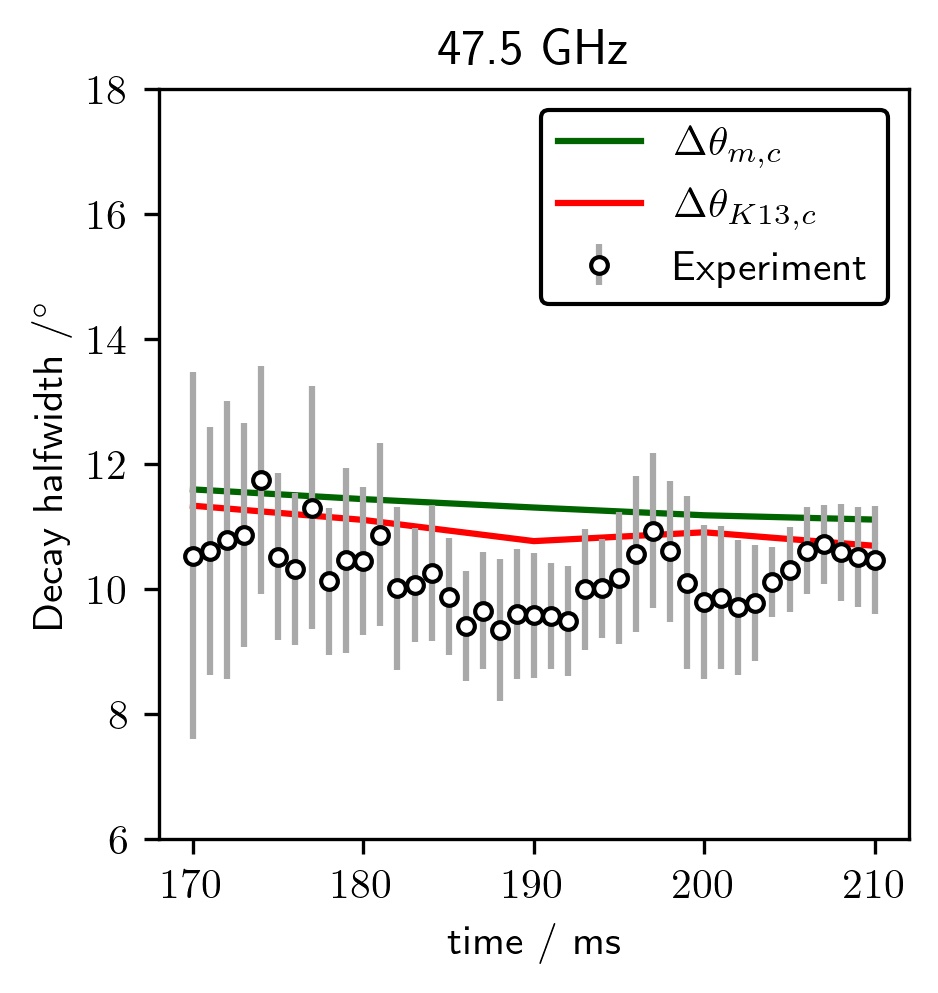}
		\end{subfigure}	
		\caption{The decay halfwidths of backscattered power as a function of time, for shot group 1: Gaussian fits of experimental data (points), the mismatch tolerance factor (green line), and the total instrumentation weight (black line). These were X-mode channels. All channels at 210 ms (left) and various times for the 47.5 GHz channel (right). In this situation, we see that the other instrumentation weights have some effect the apparent mismatch tolerance (red line), but the difference is not significant enough to be conclusively differentiated with our experimental data.}
		\label{fig:MAST_Qband}
	\end{figure*}

	Assuming a constant offset of $2.6^{\circ}$ for all Q-band channels at all times, as justified in \ref{appendix:offset}, we find that there is generally good agreement between the beam model and experimental data, see Figure \ref{fig:MAST_Qband}. The difference in mismatch tolerance calculated from $F_m$ alone compared to that with the other instrumentation effects and turbulence spectrum is still small compared to the error bars. However, the difference is larger than that of the V-band channels, see Figure \ref{fig:MAST_Vband}. The Q-band channels' cut-off locations did not vary significantly with toroidal angle, unlike the V-band's; however, the percentage change in wavenumber was larger. Hence, the wavenumber at cut-off is more important than the exact cut-off location.

	\section{Future work}
	Having a finer sweep in toroidal angle would enable one to understand the effects of the ray and beam instrumentation functions, as well as the turbulence spectrum; while this could be done with more repeated shots, small errors in exactly reproducing the plasma mean that either simultaneous measurements with an antenna array or in-shot toroidal steering would be the best approaches. For example, when the turbulence spectrum is accounted for, the beam model sometimes predicts a narrower peak in the toroidal response than the current data would suggest. Another diagnostic effect pertains to how localised is the backscattered signal to the cut-off and to what extent the signal is line integrated; having finer toroidal measurements would allow us to distinguish between these two cases. While we did calculate the expected line-integrated response with Scotty, given our data, the results were not sufficiently distinguishable from the calculated cut-off-only response. Finally, such fine measurements would require careful in-laboratory measurements of the DBS quasioptics, such that the initial conditions of the beam are accurately known. We have done our due diligence, as described in \ref{appendix:beam_parameters}, but dedicated measurements would be needed for the type of experiments proposed in this paragraph.

	\section{Conclusion} \label{section_conclusion}
	In this work, we applied the beam model of DBS \cite{Hall-Chen:beam_model_DBS:2022} via the predictive code, Scotty, to spherical-tokamak plasmas, a wider range of frequencies, and a different polarisation. Since spherical tokamaks have larger pitch angles and magnetic shear than conventional tokamaks (such as DIII-D), they provide a more comprehensive testbed for evaluating our ability to quantitatively calculate mismatch attenuation. This validation was performed at multiple times within the same shot, providing a significantly larger body of data than previously attained. In situations where the wavenumber at cut-off and cut-off location varied concurrently with changing mismatch attenuation, Scotty was able to account for the instrumentation effects and isolate the mismatch response.
	
	The findings presented in this paper paves the way for reliable and quantitative accounting of mismatch attenuation and instrumentation effects in all future DBS studies, which has two advantages: better interpretation of DBS data and loosening of the requirement that the beam wavevector has to be exactly perpendicular to the magnetic field at the cut-off. As long as the mismatch angle is not so large that the signal is below the noise floor, the beam model allows one to correct for this effect and use the backscattered signal for other studies. Quantitative correction of mismatch attenuation enables a wider range of legacy data to now be analysed, potentially yielding more physical results from existing measurements. On the other hand, our work also helps new DBS systems to be designed in situations where toroidal steering is limited or not possible, such as for burning plasmas.
	
	
	While the exponent of the turbulence spectrum might change with radial location, along with the probe beam's properties like wavenumber, we found that the variation of peak backscattered power with toroidal steering is much more sensitive to pitch angle matching than any other factor. This strong dependence of the backscattered signal on mismatch angle raises the possibility of using DBS as a method of measuring magnetic pitch angle in the core, an attractive prospect since DBS is likely able to survive the harsh environment of burning plasmas. 
	
	

	\ack {This work was partially funded by the Urban and Green Tech Office, A*STAR, Green Seed Fund C231718014. This work was partly supported by the U.S. Department of Energy under contract numbers DE-AC02-09CH11466, DE-SC0019005, DE-SC0019352, DE-SC0020649, and DE-SC0019007. The United States Government retains a non-exclusive, paid-up, irrevocable, world-wide license to publish or reproduce the published form of this manuscript, or allow others to do so, for United States Government purposes. This work has been part-funded by the EPSRC Energy Programme [grant numbers EP/W006839/1 and EP/R034737/1].  To obtain further information on the data and models underlying this paper please contact PublicationsManager@ukaea.uk. V.H. Hall-Chen's DPhil was funded by a National Science Scholarship from A*STAR, Singapore. V.H. Hall-Chen would like to thank A.M. Hall-Chen for proofreading this paper.} 
	
	

	
	\appendix

	\section{MAST DBS beam parameters} \label{appendix:beam_parameters}
	
	The beam parameters at launch are required as input for our beam-tracing code, Scotty. These parameters determine the initial $\bm{\Psi}_{w}$ at launch. In this work, we take the beam to be launched from the steering mirror's centre. The launch beam width and curvatures depend on frequency. This subsection describes and explains our choice of initial conditions.
	
	The MAST DBS system had eight Q-band channels, as well as eight V-band channels. However, for the shots studied, the highest frequency V-band channel (75 GHz) was not digitised. The most straightforward way to determine the initial beam parameters would be to simply measure them for every channel. Since it is no longer possible to measure these parameters for the original system, we proceed with the available data. The beam width as a function of distance from the lens, for the Q-band, was measured for three different frequencies: 35 GHz, 45 GHz, and 50 GHz. For the V-band, although the beam widths were measured, these measurements were not frequency resolved. As such, we estimate the beam parameters differently for the Q and V-bands. 
	
\begin{figure}
		\centering	
		\includegraphics[width=0.9\columnwidth]{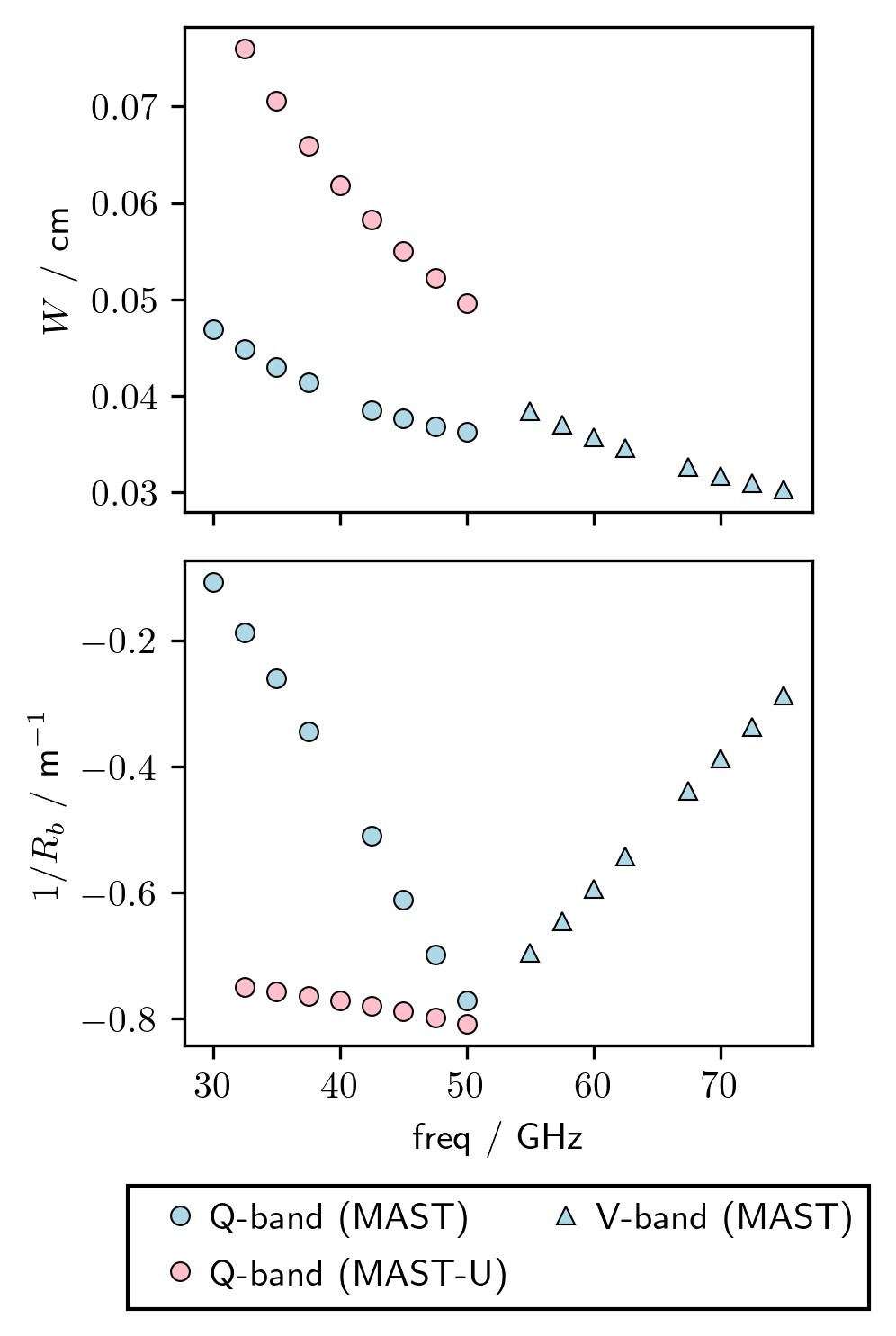}
		\caption{Beam width, $W$, and curvature, $1 / R_b$, used as initial conditions for beam tracing for the DBS systems, as a function of frequency. The MAST Q-band's initial beam was not circular, with slightly different widths and curvatures parallel and perpendicular to the polarisation. In this work, we use the width and curvature in the plane parallel to the launch polarisation.
			Determination of these parameters is discussed in \ref{appendix:beam_parameters}.}
		\label{fig:beam_params}
	\end{figure}

	To get the MAST DBS Q-band parameters for all channels, we performed full-wave simulations of the horn-lens system using the commercial software CST Studios. The beam parameters are then extracted from the far field data in the E and H planes. In \ref{appendix:beam_parameters}, we show that there is good agreement between the simulations and measured data. In this work, we use the width and curvature in the plane parallel to the launch polarisation for simplicity. Similar techniques were used for the MAST-U DBS (Q-band only), with one difference. Instead of performing full-wave simulations of the horn, we calculate the launch beam properties from the horn's specifications. As for the MAST DBS V-band, we do not know enough about the horn to perform such simulations. Consequently, for every V-band frequency, we fit the Gaussian beam evolution to the measured far-field data as shown in \ref{appendix:beam_parameters}. We summarise the beam parameters at launch in Figure \ref{fig:beam_params}.

	In this appendix, we present the methods and data used to determine the MAST DBS initial conditions required by our beam-tracing code, Scotty. At the point of writing, the MAST DBS has long since lost its original configuration. Hence, a straightforward measurement of beam properties is not possible. As such, we have seek to figure out what the launch beam widths and curvatures are, given the data we have and what we know of the horn-lens system. In the Q-band, we have three frequency resolved measurements. For each frequency, the beam widths were measured at various distances from the flat of the lens. In the V-band, similar measurements were done, but they were not frequency resolved; the measured beam width had all frequencies contributing to it at once.
	
\begin{figure} 
		\centering
		\begin{tikzpicture}[>=latex]
		
		\draw[-, very thick, black] (3, 1) -- (0,0);
		\draw[-, very thick, black] (3, -1) -- (0,0);
	
		\draw[dash dot, thin, blue] (0, 0) -- (3, 0) ;			
		
		\draw[<->, thick, blue] (3.1, 1) -- (3.1, -1) node[midway, right]{$\diameter 1.44''$} ;	
		\draw[<->, thick, blue] (0, -1.1) -- (3, -1.1) node[midway, below]{$2.7''$} ;		
		\draw[->, thick, blue] (2.5,0) arc (0:15:2.5) node[circle, fill=white!0, at start, left]{$30 ^{\circ}$};
		\draw[->, thick, blue] (2.5,0) arc (0:-15:2.5);

		\end{tikzpicture}
		\caption{Schematic of the Q-band conical horn of the MAST DBS. Lengths given in inches.} 
		\label{fig:MAST_DBS_QBand_horn}
	\end{figure}
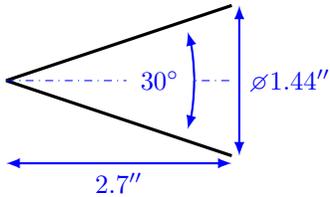

	A detailed description of the DBS setup is given in previous work \cite{Hillesheim:DBS_MAST:2015}. For each band, there was a horn antenna and a lens in front of it. The Q-band antenna was a standard smooth brass conical horn (Figure \ref{fig:MAST_DBS_QBand_horn}).
	On the other hand, the V-band antenna was a custom-made narrow-beam corrugated scalar horn, model: Quinstar QSH-V2500. Its frequency band followed the WR-15 standard (50--75GHz), the far-field full width at half maximum was $25^\circ$ in the mid-band (62.5GHz), and it was fed with a circular waveguide with a diameter of 0.165 inches. Quinstar's catalogue claims that the emission pattern is highly symmetric, and thus the far-field beam properties do not depend on the plane of measurement.
	
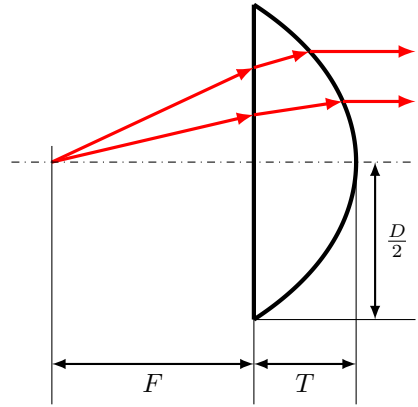
\begin{figure}
	\centering
	
	\begin{tikzpicture}[>=latex]
	
	\begin{axis}[
	tick align=outside,
	tick pos=left,
	x grid style={white!69.0196078431373!black},
	xmin=-32.5, xmax=232.5,
	xtick style={color=black},
	y grid style={white!69.0196078431373!black},
	ymin=-157.5, ymax=107.5,
	ytick style={color=black},
	axis equal image,
	axis line style={draw=none},
	tick style={draw=none},
	tick label style={draw=none},	
	xticklabels=\empty,
	yticklabels=\empty
	]
	\addplot [ultra thick, black]
	table {%
	125 -97.5
	126.786971127682 -96.3227785459
	128.572604234443 -95.1211733858474
	130.356407876571 -93.8944588981246
	132.137840888659 -92.6419002736089
	133.916309334792 -91.3627551825381
	135.691163418557 -90.0562756746328
	137.461694372619 -88.7217103283528
	139.227131351954 -87.3583066646813
	140.986638358518 -85.9653138401566
	142.739311228871 -84.5419856328543
	144.484174720338 -83.0875837336267
	146.220179735299 -81.601381353075
	147.946200727409 -80.0826671524255
	149.661033337628 -78.5307495036603
	151.363392311953 -76.9449610808722
	153.051909756549 -75.3246637808483
	154.725133789417 -73.6692539663071
	156.381527650693 -71.9781680200114
	158.019469336082 -70.2508881921571
	159.637251819425 -68.4869487170135
	161.233083931095 -66.6859421678072
	162.80509195833 -64.8475260113602
	164.351322031836 -62.9714293161073
	165.869743359657 -61.0574595589437
	167.3582523644 -59.1055094680364
	168.814677773133 -57.1155638304636
	170.236786700719 -55.0877061855109
	171.622291756746 -53.0221253169146
	172.968859193694 -50.9191214505359
	174.274118099484 -48.7791120581751
	175.535670621256 -46.6026371637646
	176.751103189228 -44.3903640453226
	177.917998690127 -42.143091225075
	179.033949519256 -39.8617516413356
	180.09657141916 -37.5474148992971
	181.103517991682 -35.2012885040068
	182.052495749349 -32.8247179876012
	182.941279552313 -30.4191858543928
	183.767728258945 -27.9863092815945
	184.529800402479 -25.5278365301919
	185.225569693319 -23.0456420394867
	185.853240137484 -20.541720199799
	186.411160556632 -18.0181778202837
	186.897838294622 -15.4772253322675
	187.311951899967 -12.9211667923388
	187.652362582923 -10.3523887729776
	187.918124260392 -7.77334825110318
	188.108492021013 -5.18655962586752
	188.222928866512 -2.59458101565782
	188.261110612975 0
	188.222928866512 2.59458101565782
	188.108492021013 5.1865596258675
	187.918124260392 7.77334825110316
	187.652362582923 10.3523887729776
	187.311951899967 12.9211667923388
	186.897838294622 15.4772253322675
	186.411160556632 18.0181778202836
	185.853240137484 20.541720199799
	185.225569693319 23.0456420394867
	184.529800402479 25.5278365301919
	183.767728258945 27.9863092815945
	182.941279552313 30.4191858543928
	182.052495749349 32.8247179876012
	181.103517991682 35.2012885040068
	180.09657141916 37.5474148992971
	179.033949519256 39.8617516413355
	177.917998690127 42.143091225075
	176.751103189228 44.3903640453226
	175.535670621256 46.6026371637645
	174.274118099484 48.7791120581751
	172.968859193694 50.9191214505359
	171.622291756746 53.0221253169146
	170.236786700719 55.0877061855109
	168.814677773133 57.1155638304635
	167.3582523644 59.1055094680364
	165.869743359657 61.0574595589436
	164.351322031836 62.9714293161074
	162.80509195833 64.8475260113602
	161.233083931095 66.6859421678072
	159.637251819425 68.4869487170135
	158.019469336082 70.250888192157
	156.381527650693 71.9781680200114
	154.725133789417 73.6692539663071
	153.051909756549 75.3246637808483
	151.363392311953 76.9449610808722
	149.661033337628 78.5307495036603
	147.946200727409 80.0826671524255
	146.220179735299 81.601381353075
	144.484174720338 83.0875837336267
	142.739311228871 84.5419856328543
	140.986638358518 85.9653138401566
	139.227131351954 87.3583066646813
	137.461694372619 88.7217103283528
	135.691163418557 90.0562756746328
	133.916309334792 91.3627551825381
	132.137840888659 92.6419002736089
	130.356407876571 93.8944588981246
	128.572604234443 95.1211733858474
	126.786971127682 96.3227785459
	125 97.5
	};
	
	\addplot [ultra thick, black]
	table {%
	125 -97.5
	125 97.5
	};		

	\draw[->, very thick, red] (axis cs:0, 0) -- (axis cs:125, 58.5);
	\draw[->, very thick, red] (axis cs:125, 58.5) -- (axis cs:159.63725181942476, 68.4869487170134);
	\draw[->, very thick, red] (axis cs:159.63725181942476, 68.4869487170134) -- (axis cs:225, 68.48694871701346);
	
	\draw[->, very thick, red] (axis cs:0, 0) -- (axis cs:125, 29.25);
	\draw[->, very thick, red] (axis cs:125, 29.25) -- (axis cs:180.09657141916017, 37.54741489929707);
	\draw[->, very thick, red] (axis cs:180.09657141916017, 37.54741489929707) -- (axis cs:225, 37.54741489929707);

	\draw[-, thin, black, dash dot] (axis cs: -25, 0) -- (axis cs: 225, 0);
	\draw[-, thin, black] (axis cs: 125, -97.5) -- (axis cs: 225, -97.5);	
	\draw[-, thin, black] (axis cs: 0, 10) -- (axis cs: 0, -150);
	\draw[-, thin, black] (axis cs: 125, 25) -- (axis cs: 125, -150);
	\draw[-, thin, black] (axis cs: 188.261110612975, 0) -- (axis cs: 188.261110612975, -150);
					
	\draw[<->, thick] (axis cs: 0,-125) -- (axis cs: 125,-125) node[midway, below]{$F$}; 
	\draw[<->, thick] (axis cs: 125,-125) -- (axis cs: 188.261110612975,-125) node[midway, below]{$T$}; 
	\draw[<->, thick] (axis cs: 200,0) -- (axis cs: 200,-97.5) node[midway, right]{$\frac{D}{2}$}; 	
					
	\end{axis}
				
	\end{tikzpicture}

	\caption{Both lenses are made of high-density polyethylene which has a refractive index of $N = 1.53$. The Q-band lens has focal length $F = 27$ cm and diameter $D = 20$ cm, while the V-band lens $F = 12.5$ cm and diameter $D = 19.5$ cm. The locus of points on the curved side $(X, Y)$ is given by equation (\ref{eq:aspheric_lens}). The flat side of the Q-band lens was $27$ cm away from the aperture of the corresponding horn, while this distance was $13.9$ cm for the V-band horn.
	} 
	\label{fig:lens}
	\end{figure}

	As for the lenses, they were aspherical hyperbolic-planar. As the name implies, one side of the lens was planar and the other was hyperbolic. The lenses were designed such that rays originating from the focal point would leave the lens parallel to the optic axis (Figure \ref{fig:lens}). We do not further discuss the design considerations in this paper. We use a fit to describe the hyperbolic side of the lens,
	\begin{equation} \label{eq:aspheric_lens}
		X = \frac{Y^2 R_l^{-1}}{1 + \left[1 - \left( 1 + k \right) Y^2 R_l^{-2}\right]^{1 / 2} } .
	\end{equation}
	Here $X$ and $Y$ are the horizontal and vertical coordinates of the hyperbolic surface, $R_l$ is the radius of curvature of the lens, and $k$ is the conic constant. The fit parameters and lens diameter and focal lengths are given in Table \ref{table:lens} and illustrated in Figure \ref{fig:lens}. We take the refractive indices for both lenses to be $N = 1.53$. The notation used in this appendix is separate from the rest of the paper.
	\begin{table} 
		\centering
		
		\begin{tabular}{ |p{3cm}||p{1.5cm}|p{1.5cm}|  }
			\hline
			Lens parameter & Q-band & V-band \\
			\hline
			$R_l / \textrm{cm}$ & $-15.5$ & $-8.78$  \\
			$k$ & $-0.594$ & $-0.588$  \\ 
			$D / \textrm{cm}$ & $20$ & $19.5$  \\ 
			$F / \textrm{cm}$ & $27$ & $12.5$  \\ [1ex] 
			\hline
		\end{tabular}
		
		\caption{Parameters for describing the hyperbolic side of the lens. Here $R_l$ is the radius of curvature of the lens in question, $k$ its conic constant, $D$ the diameter, and $F$ the focal length. We take the refractive indices for both lenses to be $N = 1.53$.}
		\label{table:lens}
	\end{table}

\begin{figure*} 
		\centering
		\includegraphics[width=\textwidth]{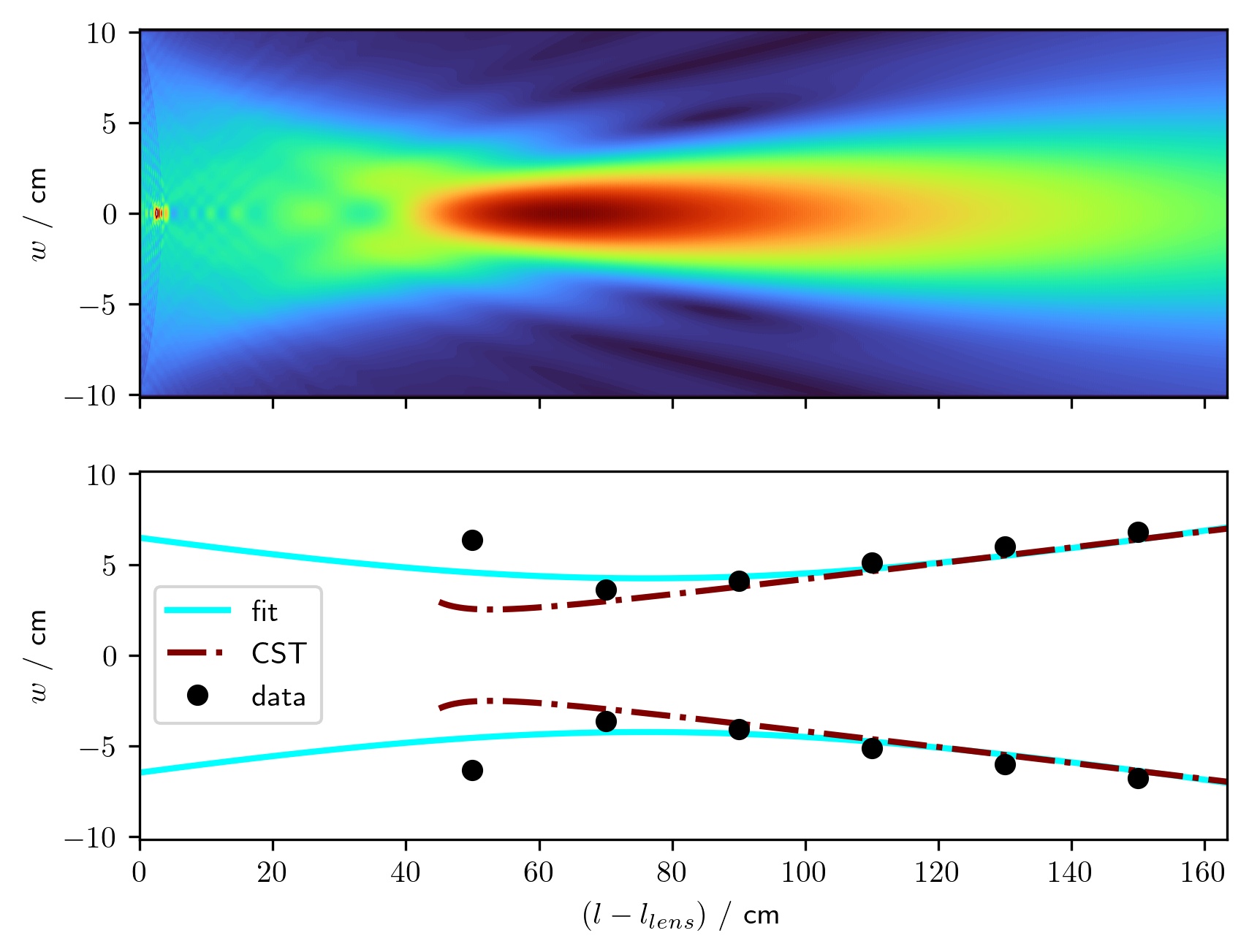}
		\caption{Magnitude of the electric field (top) and the extracted beam properties (bottom) for 35 GHz. CST simulations agree well with laboratory measurements in the far field. These CST simulations were used to find the initial beam properties for the Q-band on MAST.}
		\label{fig:CST}
	\end{figure*}

	Using this information, we run full-wave simulations of the horn-lens system to determine the Q-band beam properties. To reduce simulation time, we split this into two steps. First, the emission pattern of the Q-band antenna is simulated in CST Microwave Studio. Secondly, we feed the results into a larger simulation area, which includes the lens, as shown in Figure \ref{fig:CST} (top). We fit a Gaussian function to the beam profile, and extract the associated width. Interestingly, the beam does not reach the far field, and the profile is significantly non-Gaussian (in both simulations and in the data collected), until at least 50 cm from the lens. As such, we fit a Gaussian beam to these widths only at distances which we are confident are in the far-field Figure \ref{fig:CST} (bottom). We only fit the central lobe, as the side lobes in these simulations are larger than that measured. We find that our simulations do indeed match the data well, and so we use our simulations to determine the beam properties for the Q-band antenna.

\begin{figure} 
		\centering
		\includegraphics[width=0.99\columnwidth]{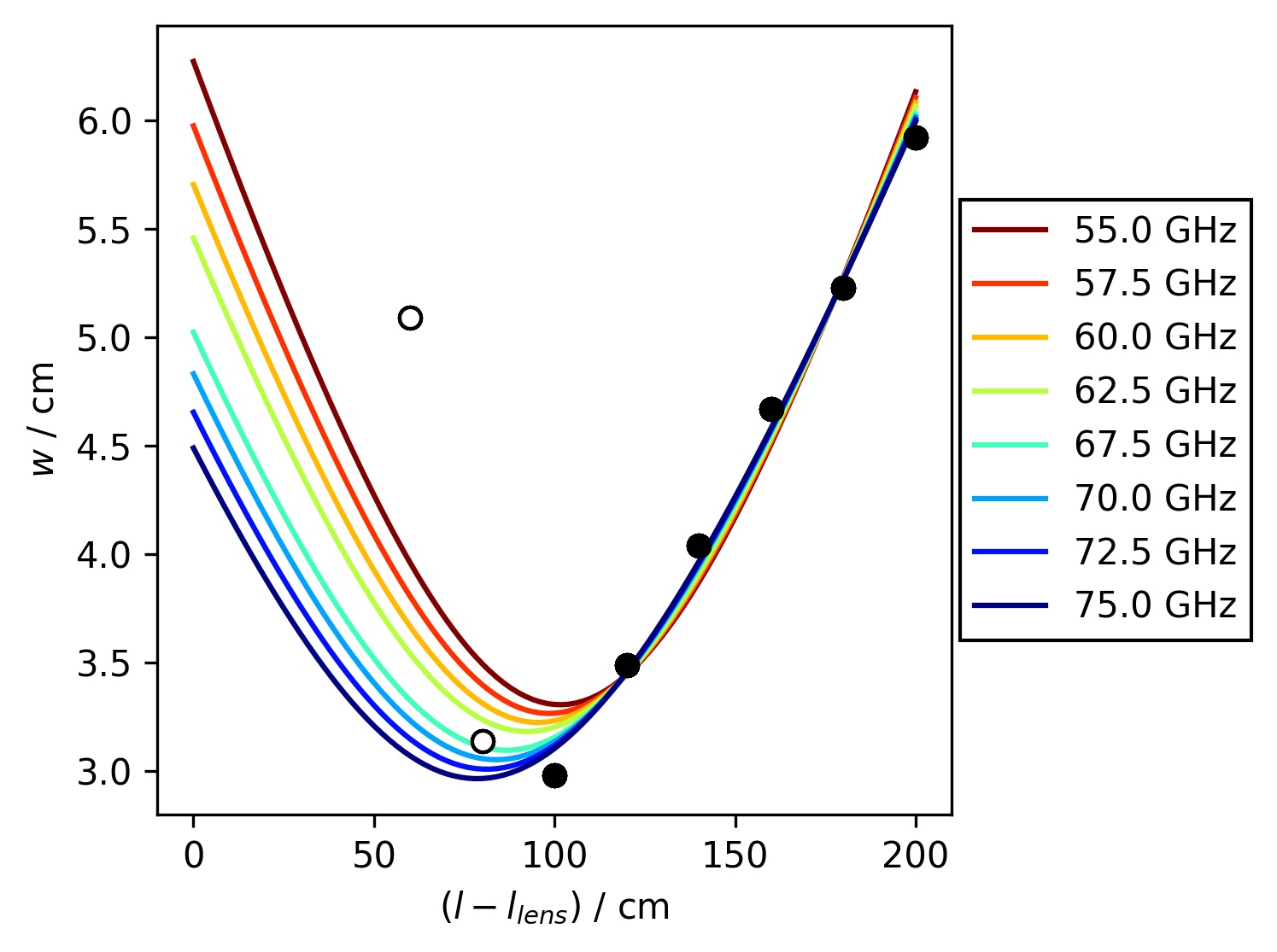}
		\caption{We use experimentally measured data, which were sufficiently far from the antenna (black-filled points), to find the beam properties for the MAST V-band frequencies. The two measurements closer to the lens (white-filled points), and hence the antenna, did not have Gaussian profiles transverse to the direction of propagation and were thus neglected.}
		\label{fig:beam_fit}
	\end{figure}

	The design of V-band horn was more complex, and we do not have enough of its details to run a meaningful simulation. Hence, our approach to finding its beam widths is more empirical than that of the Q-band. We fit all frequencies in the V-band to the far field data as shown in Figure \ref{fig:beam_fit}.
	
	The approaches described above were what we used as initial conditions for the Scotty beam-tracing simulations in this paper. We explored using other methods of obtaining these initial conditions, but they were unsatisfactory for several reasons, as described in the rest of this paragraph. We tried using horn properties, together with the thin lens approximation, to find the initial beam conditions. The result was in the right ballpark. However, as seen from Figure \ref{fig:lens}, the lenses were thick and aspherical. Hence, we calculated the appropriate ray transfer matrices for the hyperbolic lenses. We ray-traced paraxial rays through the lenses in Zemax OpticStudio 21.2. For a given lens, two paraxial rays were launched, both incident on the planar side: the first on the optical axis with a small angle of incidence, the second a small distance from the optical axis and at normal incidence. The positions and directions of both rays at the output plane were then used to calculate the ray transfer matrix. We compared the measured Q-band beam properties with that calculated from the initial beam conditions and the ray transfer matrix, and the agreement between the calculated and measured widths and curvatures was good. We repeated this for the V-band but did not manage to get decent results, likely because the initial conditions at the horn were not well-specified. Particularly, while we were able to calculate the width of the beam waist from the far-field full with at half maximum, we do not know where the waist actually is. We assumed the waist to be at the mouth of the horn.
	
	\section{Toroidal offset of the MAST Q-band system} \label{appendix:offset}
	Earlier work analysing the MAST DBS \cite{Hillesheim:DBS_MAST:2015} argued that there had to be an offset in the toroidal launch angle, for both the Q- and V-band systems. That is, there is a systematic difference between the recorded angle of the steering mirror's rotation stage and the actual angle. When varying the toroidal launch angle, we expect there to be maximum signal when the beam's wavevector is perpendicular to the magnetic field. This is a well-known result in the field, typically optimised by ray tracing. Our beam model enables the decay width of the toroidal response to be predicted, in addition to the change in the optimal toroidal angle due to other instrumentation effects. We found that the optimal toroidal angle, that is the angle where backscattered power is maximised, is only weakly affected by instrumentation effects other than mismatch. As such, we still expect there to be maximum backscattered power at the toroidal angle where the beam is fully matched at cut-off.
	
\begin{figure}
		\centering
		\includegraphics[width=0.9
		\columnwidth]{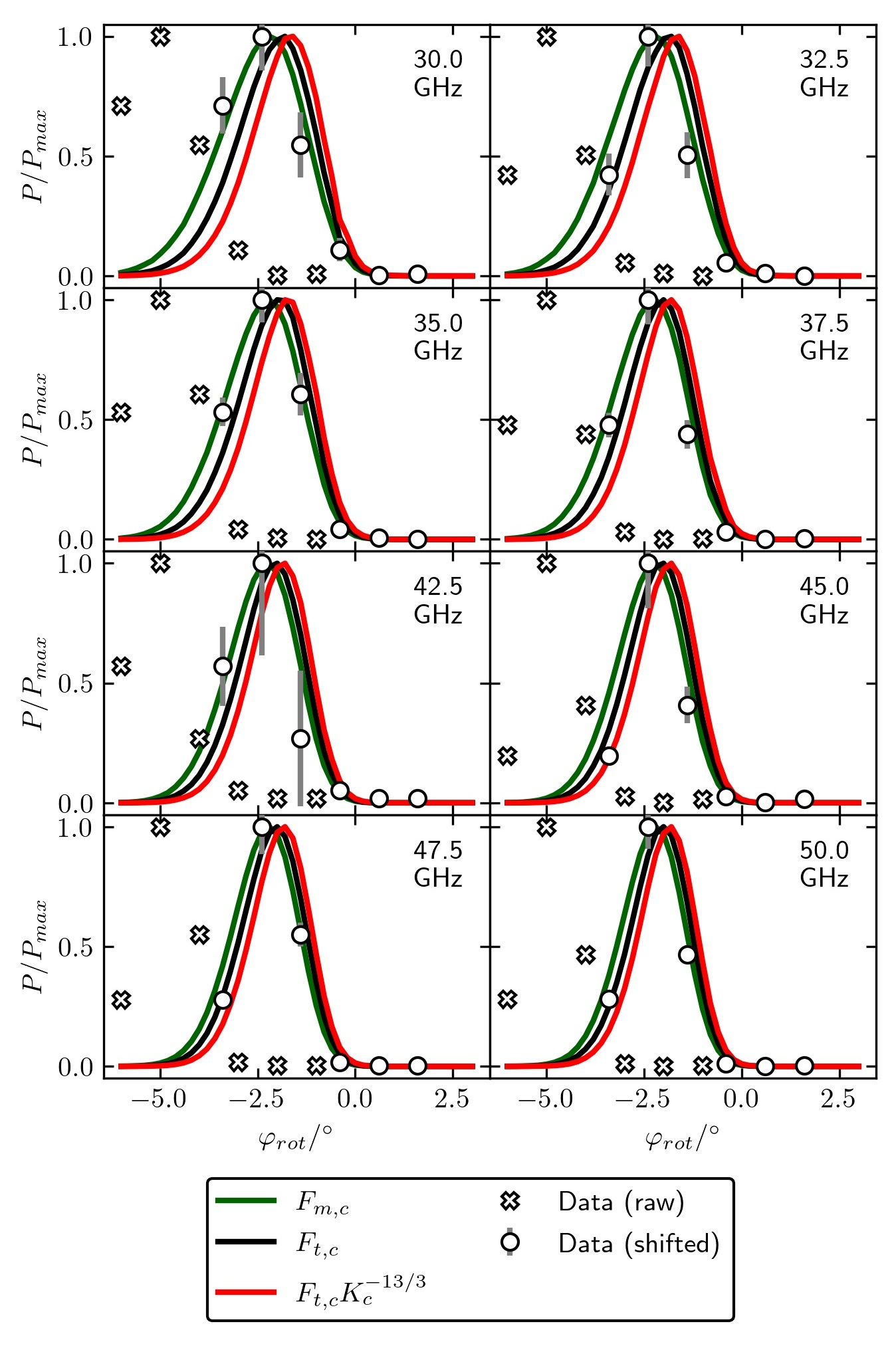} 
		\caption{Q-band data from shot group 1 at $200$ ms. Data (crosses), data assuming that there was a slight error in the launch angle (circles), model of the mismatch-attenuated backscattered signal from the cut-off (green line), model of backscattered signal (black line). Having a fixed offset in the mirror rotation angle accounts for this difference for all frequencies and all times, the latter is not shown here.
		}
		\label{fig:mismatch_Qband}
	\end{figure}

	After careful analysis of the DBS system's geometry, detailed in Section \ref{subsection:DBS_geometry}, we found that there was no offset for the V-band system.	However, for the Q-band, we find that there to be a finite offset. Since both systems use the same steering mirror, the error is not in the steering mirror's angle. Rather, we argue that there is some misalignment in the Q-band optics that is equivalent to a systematic $2.6^{\circ}$ error in the mirror angle. This accounts for most of the aforementioned discrepancy, for all Q-band channels at all times studied, see Figure \ref{fig:mismatch_Qband}. As such, in this paper, we consistently used a $2.6^{\circ}$ offset for Q-band results.

	We postulate several reasons for this offset. Since the lens had quite a short focal length, a slight $\sim 1$ mm transverse offset of the lens relative to the horn would already steer the beam enough to account for the $2.6^{\circ}$ error. Similarly, the lens might also not have been centred. A combination of lens and horn offsets could account for the angular offset seen here. Analysis of other repeated shots on MAST indicate a similarly-sized error, but the results are not as conclusive as that from shot group 1. We find that this error changes slightly, less than a degree, at different times in shot group 1, which is small enough to be attributed to the uncertainty in equilibrium reconstruction. Unfortunately, the MAST DBS was removed immediately after MAST's final campaign; experimentally-verified calibration and more detailed measurements are no longer possible.

	
	\section*{References}
	\bibliographystyle{iopart-num}
	\bibliography{references}
	
\end{document}